\documentclass[twocolumn]{aastex631}

\newcommand{\rvf}{}
\newcommand{\rvs}{}

\usepackage[T1]{fontenc}
\shorttitle{Grain Growth and Dust Segregation in WL 17}
\shortauthors{Han et al.}

\begin{document}

\title{Grain Growth and Dust Segregation Revealed by Multi-wavelength Analysis of the Class I Protostellar Disk WL 17}

\author[0000-0002-9143-1433]{Ilseung Han}
\affiliation{Department of Astronomy and Space Science, University of Science and Technology, 217 Gajeong-ro, Yuseong-gu, Daejeon 34113, Republic of Korea}
\affiliation{Korea Astronomy and Space Science Institute, 776 Daedeok-daero, Yuseong-gu, Daejeon 34055, Republic of Korea}

\author[0000-0003-4022-4132]{Woojin Kwon}
\affiliation{Department of Earth Science Education, Seoul National University, 1 Gwanak-ro, Gwanak-gu, Seoul 08826, Republic of Korea}
\affiliation{SNU Astronomy Research Center, Seoul National University, 1 Gwanak-ro, Gwanak-gu, Seoul 08826, Republic of Korea}
\correspondingauthor{Woojin Kwon}
\email{wkwon@snu.ac.kr}

\author[0000-0002-8238-7709]{Yusuke Aso}
\affiliation{Korea Astronomy and Space Science Institute, 776 Daedeok-daero, Yuseong-gu, Daejeon 34055, Republic of Korea}

\author[0000-0001-7258-770X]{Jaehan Bae}
\affiliation{Department of Astronomy, University of Florida, Gainesville, FL 32611, USA}

\author[0000-0002-9209-8708]{Patrick Sheehan}
\affiliation{National Radio Astronomy Observatory, 520 Edgemont Rd., Charlottesville, VA 22903, USA}



\begin{abstract}
    The first step toward planet formation is grain growth from (sub-)micrometer to millimeter/centimeter sizes.
    Grain growth has been reported not only in Class II protoplanetary disks but also in Class 0/I protostellar envelopes.
    However, early-stage grain growth occurring in Class 0/I stages has rarely been observed on the protostellar disk scale.
    Here we present the results from the ALMA Band 3 ($\lambda = 3.1$ mm) and 7 ($\lambda = 0.87$ mm) archival data of the Class I protostellar disk WL 17 in the $\rho$ Ophiuchus molecular cloud.
    {\rvf Disk substructures are found in both bands, but they are different:} while a central hole and a symmetric ring appear in Band 3, an off-center hole and an asymmetric ring are shown in Band 7.
    Furthermore, we obtain an asymmetric spectral index map with a low mean value of $\alpha$ $=$ 2.28 $\pm$ 0.02, suggestive of grain growth and dust segregation on the protostellar disk scale.
    Our radiative transfer modeling verifies these two features by demonstrating that 10 cm-sized large grains are symmetrically distributed, whereas 10 $\mu$m-sized small grains are asymmetrically distributed. Also, the analysis shows that the disk is expected to be massive and gravitationally unstable.
    We thus suggest a single Jupiter-mass protoplanet formed by gravitational instability as the origin of the ring-like structure, grain growth, and dust segregation identified in WL 17.
\end{abstract}

\keywords{Protostars (1302), Circumstellar disks (235), Circumstellar dust (236), Circumstellar grains (239)}



\section{Introduction}
\label{sec:introduction}

Protoplanetary disks, circumstellar disks of the so-called Class II young stellar objects (YSOs), are the natal place of planets.
However, it is {\rvf unknown} when planet formation begins.
Up to now, the youngest protoplanets have been identified only in intermediate-aged protoplanetary disks through optical to (sub-)mm observations: a Jupiter-mass planet around AS 209 \citep[1$-$2 Myr;][]{2022ApJ...934L..20B}
and AB Aur \citep[4 Myr;][]{2022NatAs.tmp...76C}, and two super-Jovian-mass planets around PDS 70 \citep[5 Myr;][]{2018A&A...617A..44K, 2019NatAs...3..749H}, and a potential Neptune-mass planet around TW Hya \citep[10 Myr;][]{2019ApJ...878L...8T}.
On the other hand, exoplanet surveys have shown that protoplanetary disks do not have enough mass to form planets, implying that planets should form before the Class II stage \citep[e.g.,][]{2010MNRAS.407.1981G, 2014MNRAS.445.3315N, 2018A&A...618L...3M}.
Theoretical studies support this idea by demonstrating that planets, particularly Jupiter-mass gas giants, can rapidly form by gravitational instability (GI) before the Class II stage, i.e., within $\sim$1 Myr \citep[e.g.,][and references therein]{1997Sci...276.1836B, 1998ApJ...503..923B, 2002Sci...298.1756M, 2004ApJ...609.1045M, 2007prpl.conf..607D}.
Recently, systematic observational studies in (sub-)mm wavelengths have also suggested the possibility of planet formation in the Class 0/I protostellar stages \citep[e.g.,][]{2018ApJS..238...19T, 2020A&A...640A..19T, 2019ApJ...875L...9W, 2020ApJ...890..130T}.
Indeed, a systematic study toward young circumstellar disks has also been carrying out \citep[e.g.,][]{2023ApJ...951....8O}.
Now planet formation is expected to begin in the
early protostellar stages.

Grain sizes are estimated by a dust opacity spectral index ($\beta$) in (sub-)mm wavelengths \citep[e.g.,][]{1993Icar..106...20M, 2001ApJ...553..321D, 2006ApJ...636.1114D, 2009ApJ...696..841K, 2010A&A...512A..15R}.
$\beta$ {\rvf is usually} derived from a spectral index ($\alpha$) in the optically thin and Rayleigh-Jean's approximation case as $\alpha$ $=$ $\beta$ $+$ 2 (see also Section \ref{sec:spectral_index}).
Based on theoretical studies, the index is also related to other dust properties, such as shape, composition, and porosity \citep[e.g.,][]{1994ApJ...421..615P, 2001ApJ...553..321D, 2014A&A...568A..42K}, but above all, it is highly sensitive to the size: specifically, $\beta$ $\lesssim$ 1.0 at a 1-mm wavelength indicates a grain size larger than 3 mm \citep{2006ApJ...636.1114D}.
Indeed, many observational studies in (sub-)mm wavelengths have so far reported the presence of mm/cm-sized large grains in
protoplanetary disks by showing such low mean $\beta$ values \citep[e.g.,][]{1991ApJ...381..250B, 2006A&A...446..211R, 2007ApJ...659..705A, 2007A&A...462..211L, 2010A&A...521A..66R, 2010A&A...512A..15R, 2011ApJ...741....3K, 2015ApJ...808..102K, 2021MNRAS.506.5117T}, compared with the interstellar medium (ISM) consisting of submicrometer-sized small grains ($\beta_{\rm ISM}$ $\simeq$ 1.7; e.g., \citealt{1999ApJ...524..867F, 2001ApJ...554..778L, 2014A&A...566A..55P, 2014A&A...571A..11P}).

Grain growth occurs even in the very early evolutionary stages of YSOs.
Considering the angular resolutions of previous interferometric observations, grain growth ($\beta$ $\lesssim$ 1.0) has mainly been reported in the inner envelopes \citep[e.g.,][]{2007ApJ...659..479J, 2009ApJ...696..841K, 2015ApJ...808..102K, 2012ApJ...756..168C, 2014A&A...567A..32M, 2019A&A...632A...5G} and the outer disks \citep[e.g.,][]{2013ApJ...771...48T} around Class 0/I protostars.
Recently, higher-resolution observations with ALMA and the Karl G. Jansky Very Large Array (VLA) allow us to probe grain growth in Class I protostellar disks in more detail.
For example, the presence of 10 cm-sized large grains was suggested in the outer region of the edge-on disk CB 26, although such rapid grain growth could not be probed in the inner region due to high optical depth \citep[][]{2021A&A...646A..18Z}.
Grain growth even
to cm size was reported in the less inclined disk EC 53 ($i$ $=$ 34.8$^{\circ}$; \citealt{2020ApJ...889...20L}).
In addition to the $\beta$ analysis, \citet{2018NatAs...2..646H} suggested rapid grain growth to mm size implied by a lack of CO isotopologue emissions in the inner region ($R_{\rm disk}$ $\lesssim$ 30 au) of TMC-1A.
Grain growth occurs in disks non-uniformly.
It has been reported that $\beta$ values of central regions in YSO envelopes and disks are smaller than those of outer regions \citep[e.g.,][]{2009ApJ...696..841K, 2011A&A...529A.105G, 2012ApJ...760L..17P, 2015ApJ...813...41P, 2016A&A...588A..53T}.
In addition, $\beta$ values show a dependence on substructures of protoplanetary disks, which have been detected by high-resolution ALMA observations \citep[e.g.,][]{2016ApJ...829L..35T, 2022ApJ...928...49T, 2019ApJ...881..159M, 2021A&A...648A..33M, 2019ApJ...883...71C, 2020ApJ...898...36L, 2021ApJS..257...14S}.
Particularly in the ring region, $\beta$ is smaller than 1.0, indicating that grains grown to mm/cm sizes are concentrated in this region.
Such ring-like substructures are recently revealed in Class 0/I protostellar disks as well \citep[e.g.,][]{2017ApJ...840L..12S, 2018ApJ...857...18S, 2020A&A...634L..12D, 2020ApJ...895L...2N, 2020NatAs...4..142L, 2020Natur.586..228S, 2020ApJ...902..141S, 2022ApJ...934...95S, 2020ApJ...904L...6A}, where grain growth may be enhanced.

Within these disk substructures, particularly consisting of inner holes and outer rings, the spatial distribution of dust grains is different depending on grain size.
For example, previous near-infrared (NIR) and (sub-)mm observations toward protoplanetary disks have shown that in shorter wavelengths, the size of the observed holes becomes smaller \citep[e.g.,][]{2012ApJ...758L..19H, 2015ApJ...799...43H, 2012ApJ...760..111D, 2017ApJ...836..201D, 2014ApJ...791...42Z, 2015A&A...584L...4P, 2018MNRAS.475L..62H, 2019A&A...625A.118K},
and the dust scale height larger \citep[e.g,][]{2019A&A...624A...7V, 2020A&A...642A.164V}.
It means that $\mu$m-sized small grains are more widely distributed than mm/cm-sized large grains in the radial direction, and they can be lifted up to the disk surface, for example, due to turbulence in the vertical direction.
Furthermore, within the (sub-)mm wavelength regime, the spatial distribution between
mm- and cm-sized grains
is slightly different in both radial and vertical directions.
According to ALMA multi-wavelength observations toward protoplanetary disks of Class II YSOs, the width of the rings is narrower in longer wavelengths \citep[e.g.,][]{2017ApJ...839...99P, 2019ApJ...878...16P}, and the dust scale height is smaller in longer wavelengths \citep[e.g.,][]{2020A&A...642A.164V, 2022ApJ...930...11V}.
It indicates that larger grains are more concentrated in the ring and more settled down toward the disk midplane.
However, whether or not dust segregation happens in Class 0/I protostellar disks is still unclear.

In this paper, we present ALMA Band 3 and 7 archival data of the Class I protostellar disk WL 17 to investigate the size and spatial distribution of dust grains on the protostellar disk scale.
WL 17 is an M3-type protostar \citep{2010ApJS..188...75M} and is located in the L1688 region of the $\rho$ Ophiuchus molecular cloud \citep[$d$ = 137 pc;][]{2017ApJ...834..141O}.
According to previous observations in IR and (sub-)mm wavelengths, it has been classified as a late Class I protostar with an age of $\lesssim$0.7 Myr \citep[e.g.,][]{2009ApJ...692..973E, 2009ApJS..181..321E, 2015ApJS..220...11D}, indicating that its protostellar envelope is nearly dissipated \citep{2009ApJ...692..973E, 2009A&A...498..167V}.
A small disk around the protostar has been revealed by multiple ALMA dust continuum observations \citep[][]{2017ApJ...840L..12S, 2019ApJS..245....2S, 2019MNRAS.482..698C, 2021ApJ...922..150G}.
Particularly, despite its compact size, the Class I protostellar disk clearly shows substructures consisting of a large central hole ($R_{\rm hole}$ = 12 au) and a horseshoe-like narrow ring ($\sigma_{\rm ring}$ = 11 au; $R_{\rm disk}$ = 23 au) in the high-resolution Band 3 image \citep{2017ApJ...840L..12S}, which may imply the possibility of grain growth within the ring, {\rvf similar to structured} protoplanetary disks.
Furthermore, \citet{2021ApJ...922..150G} recently reported that the disk substructures are also resolved in Band 7, and the disk is geometrically flared based on the marginally resolved Band 6 image.
The disk is relatively more massive than other Class I disks in the $\rho$ Ophiuchus molecular cloud ($M_{\rm dust}$ $=$ 13$-$32 $M_{\earth}$; e.g., \citealt{2019ApJ...875L...9W, 2019ApJS..245....2S}).
It implies that planets are more likely to form in this massive disk because the planet formation efficiency is predicted to increase in disks with more available material \citep[e.g.,][]{2013ApJ...771..129A}.
For these reasons, the clearly-structured disk WL 17 is one of the best targets for studying grain growth and dust segregation on the protostellar disk scale through multi-wavelength analysis.

This paper is organized as follows.
In Section \ref{sec:observations}, we describe the observational details of the ALMA Band 3 and 7 archival data, and the data reduction and imaging procedure.
In Section \ref{sec:observational_results}, we present ALMA Band 3 and 7 dust continuum images and the spectral index map.
In Section \ref{sec:modeling_analysis}, to investigate the size and spatial distribution of dust grains in the disk, we perform radiative transfer modeling and analyze the modeling results.
In Section \ref{sec:discussion}, we discuss these modeling results in the context of planet formation.
Lastly, our conclusions are summarized in Section \ref{sec:conclusions}.

\section{Observations and Data Reduction}
\label{sec:observations}

\begin{deluxetable*}{cccccccccc}
    \label{tab:tab1}
    \tablenum{1}
    \tablecaption{Summary of ALMA Observations}
    \tablehead{
        \colhead{Band} &
        \colhead{Date} &
        \colhead{Freq. Range} &
        \colhead{Antennas} &
        \colhead{Config.} &
        \colhead{Baselines} &
        \colhead{On-source Time} &
        \colhead{} &
        \colhead{Calibrators} &
        \colhead{} \\
        \colhead{} &
        \colhead{} &
        \colhead{(GHz)} &
        \colhead{} &
        \colhead{} &
        \colhead{(m)} &
        \colhead{(minutes)} &
        \colhead{Flux} &
        \colhead{Bandpass} &
        \colhead{Phase}}
    \startdata
        3 & 2015 Oct 31 &  89.50-105.49 & 38 & C-8/7 & 85-16196 & 2.82 & J1517-2422 & J1517-2422 & J1625-2527 \\ 
          & 2015 Nov 26 &  89.50-105.49 & 37 & C-8/7 & 68-14321 & 2.82 & J1517-2422 & J1517-2422 & J1625-2527 \\ 
          & 2016 Apr 17 &  89.49-105.48 & 40 & C-2/3 & 15-601   & 0.97 & J1733-1304 & J1427-4206 & J1625-2527 \\ 
        7 & 2016 May 19 & 342.01-357.24 & 40 & C-3   & 15-640   & 0.40 & J1517-2422 & J1517-2422 & J1625-2527 \\ 
          & 2016 Sep 11 & 342.01-357.24 & 37 & C-6   & 15-3144  & 0.91 & J1517-2422 & J1517-2422 & J1625-2527    
    \enddata
    \tablecomments{In this paper, all the Band 3 data were used, but we used only the extended configuration (C-6) data in Band 7.}
\end{deluxetable*}

We used the ALMA archival data of the Class I protostellar disk WL 17 taken in Band 3 and 7 during Cycle 3 (2015.1.00761.S; PI: Patrick Sheehan).
As shown in Table \ref{tab:tab1}, the Band 3 observations were made in two configurations (C-8/7 and C-2/3) from 2015 October 31 to 2016 April 17.
Each configuration had the same spectral setup using 4 spectral windows centered at 90.495, 92.432, 102.495, and 104.495 GHz with a 2-GHz bandwidth.
In the extended configuration (C-8/7), two execution blocks were taken with the same calibrators: J1517$-$2422 for flux and bandpass calibration and J1625$-$2527 for phase calibration.
The flux densities of J1517$-$2422 were set to be 2.256 Jy at 97.479 GHz with a spectral index of $-$0.234 and 2.555 Jy with an index of $-$0.300 for the two execution blocks.
The flux densities of J1625$-$2527 were bootstrapped as 0.795, 0.784, 0.736, and 0.725 Jy for individual spectral windows of the first execution block.
They were 0.821, 0.807, 0.749, and 0.739 Jy for the second execution block.
The numbers of antennas used for these two execution blocks were 38 and 37, respectively.
In the compact configuration (C-2/3), J1733$-$1304 was a flux calibrator, while J1427$-$4206 was a bandpass calibrator.
The flux density of J1733$-$1304 was set to 3.279 Jy at 90.495 GHz with a spectral index of $-$0.562.
Like the extended configuration, J1625$-$2527 was employed as a phase calibrator, and its flux densities were bootstrapped as 0.689, 0.680, 0.633, and 0.626 Jy in individual spectral windows.
The number of antennas used was 40.

The Band 7 observations were made in two configurations (C-3 and C-6) from 2016 May 19 to 2016 September 11.
The same calibrators as the Band 3 extended configuration observations were used in both configurations of the Band 7 observations: J1517$-$2422 for flux and bandpass calibration and J1625$-$2527 for phase calibration.
The spectral setup in the compact configuration (C-3) had 5 spectral windows centered at 343.018, 344.219, 345.358, 354.524, and 356.269 GHz with bandwidths of 2 GHz, 117.188, 117.188, 234.375 MHz, and 2 GHz, respectively.
The flux density of J1517$-$2422 was set to 1.914 Jy at 348.678 GHz with a spectral index of $-$0.265.
The flux densities of J1625$-$2527 were calculated to be 0.252 Jy at 343.018 GHz and 0.247 Jy at 356.269 GHz for the wide 2-GHz bandwidths.
The number of antennas for this configuration was 40.
In the extended configuration (C-6), the spectral setup likewise had 5 spectral windows with the same bandwidths as the compact configuration.
These spectral windows were centered at 342.978, 344.178, 345.318, 354.483, and 356.227 GHz, which are slightly different from those in the compact configuration due to the Doppler shift effect by the rotation and revolution of Earth.
The flux density of J1517$-$2422 was set to be 2.800 Jy at 342.978 GHz with a spectral index of $-$0.200.
The flux densities of J1625$-$2527 were calculated to be 0.288 Jy at 342.978 GHz and 0.282 Jy at 356.227 GHz for the wide 2-GHz bandwidths.
The number of antennas was 37.
Details of the observations are summarized in Table \ref{tab:tab1}.
In addition, general descriptions of the observations in both Band 3 and 7 are found in \citet{2017ApJ...840L..12S, 2018ApJ...857...18S} and \citet{2021ApJ...922..150G}.

The ALMA archival data were calibrated with CASA \citep{2022PASP..134k4501C} of the versions utilized in individual data reduction scripts:
CASA 4.5.0 and 4.5.3 for the Band 3 extended and compact configuration data (C-8/7 and C-2/3) and CASA 4.6.0 and 4.7.0 for the Band 7 compact and extended configuration data (C-3 and C-6), respectively.
Imaging and analysis for the Band 3 and 7 data were performed with CASA 5.4.0.

Because the Band 3 observations were spanned over about a half year as summarized in Table \ref{tab:tab1}, we considered the proper motion of WL 17.
For the proper motion correction, before combining all the execution blocks to make the final image, we first imaged individual execution blocks separately using Briggs weighting with a robust parameter of 0.5 and then compared their disk centers.
To measure the disk center, we fit an elliptical Gaussian to each image using the CASA task \textit{imfit}.
Note that the difference of the disk centers between the first two execution blocks, which were taken in the same extended configuration (C-8/7; Table \ref{tab:tab1}), was negligible due to the short time interval.
In the combined data of the extended configuration, the deconvolved Gaussian center was finally obtained as $\alpha$(J2000) = 16$^{\rm h}$27$^{\rm m}$06$\fs$77 and $\delta$(J2000) = $-$24$^\circ$38$\arcmin$15$\farcs$44.
The Gaussian center was adopted as a common disk center, and also, it was assigned as a phase center using the CASA tasks \textit{fixvis}.
In contrast, there is an obvious proper motion between the extended and compact configuration images.
To correct the proper motion, using the CASA tasks \textit{imfit}, \textit{fixvis}, and \textit{fixplanets}, the measured disk center of the compact configuration image (the third execution block in Table \ref{tab:tab1}) was shifted toward the common disk center, and also the shifted position was set as the phase center.
{\rvf The offset of the Band 3 compact configuration data from the Band 3 extended configuration data (i.e., the common disk center) is ($-$42.75 mas, 5.12 mas). The offset of the Band 7 extended configuration data from the common disk center is ($-$11.25 mas, $-$20.98 mas). Note that the proper motion reported by \citet{2017A&A...597A..90D} is ($-$10.0 $\pm$ 0.5 mas yr$^{-1}$, $-$27.9 $\pm$ 0.4 mas yr$^{-1}$), which is comparable to the offset of the Band 7 extended data from the common disk center data taken with an interval of about a year. The offset of the Band 3 compact configuration data is largely different from the corresponding proper motion value, but it is understandable considering the limited angular resolution.}
After combining all the extended and compact configuration data, we tried self-calibration as well but did not achieve a significant improvement, perhaps due to a low original S/N of 21.
We did not, therefore, include self-calibration in the Band 3 imaging.
The final image was made using Briggs weighting with a robust parameter of 0.5, which provided the best compromise regarding both angular resolution and sensitivity.
The Band 3 image has a synthesized beam of 0.074$\arcsec$ $\times$ 0.060$\arcsec$ (PA = 78$^\circ$) and a sensitivity of 33 $\mu$Jy beam$^{-1}$.
In addition, elliptical tapering (2.0M$\lambda$ $\times$ 1.5M$\lambda$, PA = 80$^{\circ}$) was employed to achieve a comparable synthesized beam size to a Band 7 image.
The tapered image has a synthesized beam of 0.108$\arcsec$ $\times$ 0.103$\arcsec$ (PA = 67$^{\circ}$) and a sensitivity of 34 $\mu$Jy beam$^{-1}$.

The Band 7 observations were also carried out in two configurations (C-3 and C-6; Table \ref{tab:tab1}).
In these two configurations, each execution block was calibrated and imaged separately using Briggs weighting with a robust parameter of 0.5.
We found that the Band 7 image made by the compact configuration data does not resolve any substructures because of a limited angular resolution of 0.331$\arcsec$ $\times$ 0.148$\arcsec$, which is larger than the entire disk size ($R_{\rm disk}$ $\lesssim$ 0.2$\arcsec$; \citealt{2017ApJ...840L..12S}).
Also, we verified that the total flux does not change significantly when combining the compact configuration data with the extended configuration: 125.9 $\pm$ 0.3 mJy for the combined data \citep{2021ApJ...922..150G} and 125.8 $\pm$ 1.36 mJy for the extended configuration only.
For the Band 7 imaging, thus, we used only the extended configuration data in order to focus on disk substructures.
{\rvf {\rvs On the other hand, the flux difference} within the 3$\sigma$ area is relatively large in Band 3 ($\sim$7$\%$): 6.84 $\pm$ 0.17 and 7.32 $\pm$ 0.14 mJy in the extended-only and combined configurations, respectively. {\rvs In Band 3 we decided to use the combined data to avoid the flux filtering issue and to} have a large {\it uv} coverage for beam matching.}
The same imaging procedure was applied to the Band 7 data set.
Using the CASA tasks \textit{imfit}, \textit{fixvis}, and \textit{fixplanets}, the measured disk center was shifted toward the common disk center, which was determined in the Band 3 extended configuration image as described above, and was also assigned as the phase center.
For the same reasons with the Band 3 imaging, we did not perform self-calibration in Band 7 either.
The final image was made using Briggs weighting with a robust parameter of 0.5, which was the best compromise between angular resolution and sensitivity.
The Band 7 image has a synthesized beam of 0.107$\arcsec$ $\times$ 0.104$\arcsec$ (PA = $-$37$^{\circ}$) and a sensitivity of 0.4 mJy beam$^{-1}$.

\section{Observational Results}
\label{sec:observational_results}

    \subsection{Band 3 Continuum}
    \label{sec:band3}


    \begin{figure}
        \centering
        \includegraphics[scale=0.6]{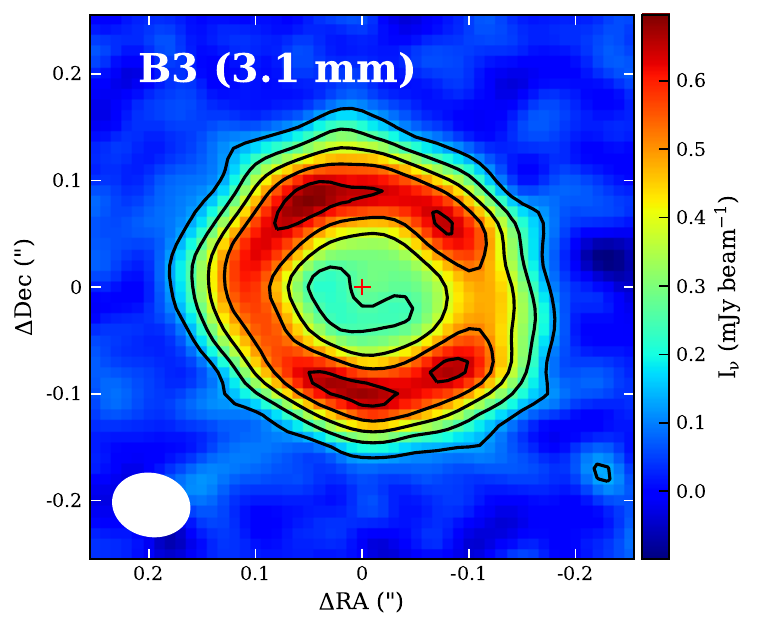}
        \caption{ALMA Band 3 (3.1 mm; 97.5 GHz) {\rvf continuum image} of the Class I protostellar disk WL 17. Contour levels are $\{$4, 8, 12, 16, 20$\}$ $\times$ $\sigma_{\rm B3}$, where $\sigma_{\rm B3}$ corresponds to 33 $\mu$Jy beam$^{-1}$. Particularly, the non-circular 8$\sigma_{\rm B3}$ contour implies the weak emission in the central hole, which was previously reported in \citet{2017ApJ...840L..12S}. The synthesized beam size shown at the lower left is 0.074$\arcsec$ $\times$ 0.060$\arcsec$ with PA $=$ 74$^{\circ}$.
        The red cross indicates the position of the protostar.}
        \label{fig:fig1}
    \end{figure}

    Figure \ref{fig:fig1} shows a Band 3 continuum image of the Class I protostellar disk in WL 17.
    Because we combined all the compact and extended configuration data listed in Table \ref{tab:tab1}, this image has a slightly lower angular resolution (0.074$\arcsec$ $\times$ 0.060$\arcsec$) than the image (0.06$\arcsec$ $\times$ 0.05$\arcsec$) presented by \citet{2017ApJ...840L..12S} that used only the extended configuration data (the first two execution blocks in Table \ref{tab:tab1}).
    Nevertheless, Figure \ref{fig:fig1} clearly reveals disk substructures: a central hole and a horseshoe-like ring around the hole.
    These substructures are consistent with those reported by \citet{2017ApJ...840L..12S} and \citet{2021ApJ...922..150G}.
    The hole has a radius of $\sim$0.06$\arcsec$ (8 au), and the ring has a width of $\sim$0.08$\arcsec$ (11 au).
    These values will be measured more specifically through radiative transfer modeling in Section \ref{sec:modeling_analysis}.
    The ring has a nearly symmetric shape but a marginally asymmetric brightness distribution in the azimuthal direction, showing a maximum intensity of 0.696 mJy beam$^{-1}$ at PA = 32$^{\circ}$ and a minimum intensity of 0.477 mJy beam$^{-1}$ at PA = 270$^{\circ}$.
    In the central hole, there is a weak emission above the 8$\sigma_{\rm B3}$ level, which was previously discovered by \citet{2017ApJ...840L..12S}, but the contrast with the background emission inside the hole is not significant, which is about 2$\sigma_{\rm B3}$.
    The total flux within the 5$\sigma_{\rm B3}$-contour region with a radius of $\sim$0.17$\arcsec$ (23 au) is measured to be 6.82 $\pm$ 0.13 mJy.
    In addition, Figure \ref{fig:fig2}a shows a tapered Band 3 image with an angular resolution of 0.108$\arcsec$ $\times$ 0.103$\arcsec$ to achieve a comparable synthesized beam size to a Band 7 image, which will be introduced in the following subsection.

    The geometry of the disk is consistent with previous studies.
    To measure the geometry, we fit an elliptical Gaussian to the high-resolution Band 3 image (Figure 1) using the CASA task \textit{imfit}.
    We obtain that the deconvolved Gaussian has an FWHM of 0.272$\arcsec$ $\pm$ 0.012$\arcsec$ $\times$ 0.235$\arcsec$ $\pm$ 0.010$\arcsec$ (37 au $\times$ 32 au) and a position angle of 58$^{\circ}$ $\pm$ 14$^{\circ}$.
    Its inclination angle is also estimated to be 30$\arcdeg^{+7\arcdeg}_{-11\arcdeg}$ from the major axis and minor axis values of the FWHM.
    This FWHM is comparable to other FWHM values obtained from previous ALMA Band 6 continuum observations \citep{2019MNRAS.482..698C, 2019ApJS..245....2S}.
    \citet{2017ApJ...840L..12S} obtained the inclination angle of 28$^{\circ}$ and the position angle of 82.4$^{\circ}$ through radiative transfer modeling only with the ALMA Band 3 extended configuration data (the first two execution blocks in Table \ref{tab:tab1}).
    Recently, \citet{2021ApJ...922..150G} estimated the inclination angle as 31.2$^{\circ}$ and the position angle as 56$^{\circ}$ through visibility modeling, using all the Band 3 data sets listed in Table \ref{tab:tab1}.
    Furthermore, according to \citet{2013A&A...556A..76V}, the $^{12}$CO (3$-$2) outflow was observed to extend in the northwest-southeast direction by the James Clerk Maxwell Telescope (JCMT).
    The outflow was measured to be inclined by 50$^{\circ}$ from the line of sight.

    Assuming that thermal continuum emission originates from isothermal dust grains and is optically thin in (sub-)mm wavelengths, a dust mass can be measured from a total flux density as follows \citep{1983QJRAS..24..267H}:
    \begin{equation}
        \label{eq:eq1}
        M_{\rm dust} = \frac{F_{\nu}d^2}{{\kappa}_{\nu}B_{\nu}(T_{\rm dust})},
    \end{equation}
    where $F_{\nu}$ is the total flux density at the frequency $\nu$, $d$ is the distance, ${\kappa}_{\nu}$ is the dust mass absorption coefficient (so-called dust opacity) at the frequency $\nu$, and $B_{\nu}(T_{\rm dust})$ is the Planck function at the dust temperature $T_{\rm dust}$.
    The total flux density measured within the 5$\sigma_{\rm B3}$-contour region is 6.82 mJy (Figure 1).
    As introduced in Section 1, the distance is 137 pc, which is the same as that used in \citet{2017ApJ...840L..12S}.
    Note that this value is the mean distance to the L1688 region in the $\rho$ Ophiuchus molecular cloud \citep{2017ApJ...834..141O}.
    The dust opacity at a central frequency of 97.5 GHz is adopted to be 0.975 cm$^2$ g$^{-1}$, which was calculated from the equation in \citet{1990AJ.....99..924B}: $\kappa_{\nu}$ = 10 ($\nu$ / 1 THz)$^\beta$ and $\beta$ = 1.
    In addition, this widely-used opacity is comparable to the opacity with the maximum grain size ($a_{\rm max}$) of 1 mm calculated by several previous studies \citep[See also Section \ref{sec:modeling_setup}; e.g.,][]{2009ApJ...700.1502A, 2011ApJ...732...42A, 2018ApJ...869L..45B, 2019MNRAS.486.3907P}.
    Regarding dust temperature, in multiple previous observations toward the $\rho$ Ophiuchus molecular cloud, it was assumed to be uniformly 20 K for calculating dust masses of the complete observed disk sample, including WL 17 \citep[][]{2007ApJ...671.1800A, 2019ApJ...875L...9W, 2019ApJS..245....2S}.
    Given the wide range of physical properties, such as bolometric luminosity ($L_{\rm bol}$), for protostars \citep[e.g.,][]{2015ApJS..220...11D}, several disk surveys have recently adopted various mean dust temperatures adjusted for each protostar \citep[e.g.,][]{2020ApJ...890..130T, 2021ApJ...913..149E}.
    Also, for WL 17, we confirmed that the mean dust temperature in the ring, where most grains are concentrated, is estimated to be 30 K by assuming the radiative equilibrium that will be discussed in Section \ref{sec:modeling_analysis}.
    Consequently, adopting $T_{\rm dust}$ = 30 K, the dust mass of the WL 17 disk is obtained to be 26 $M_{\oplus}$ in Band 3.

    \subsection{Band 7 Continuum}
    \label{sec:band7}

    \begin{figure*}
        \centering
        \includegraphics[scale=0.8]{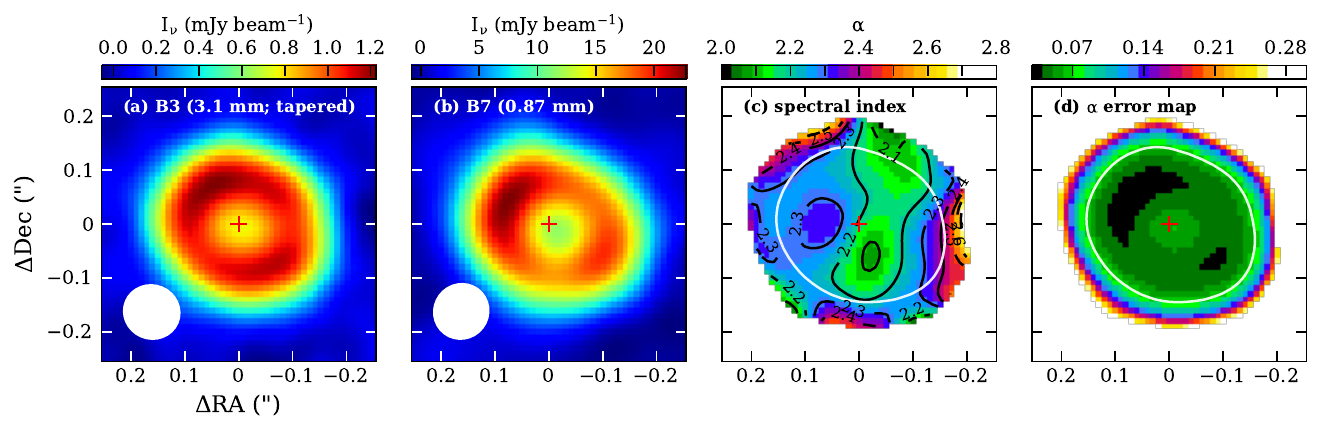}
        \caption{ALMA {\rvf images} of WL 17. (a) Tapered Band 3 dust continuum image with the synthesized beam of 0.108$\arcsec$ $\times$ 0.103$\arcsec$ (PA $=$ 67$^{\circ}$) with the sensitivity of 34 $\mu$Jy beam$^{-1}$. The red cross indicates the position of the protostar. The original Band 3 image is shown in Figure \ref{fig:fig1}. (b) Band 7 dust continuum image with the synthesized beam of 0.107$\arcsec$ $\times$ 0.104$\arcsec$ (PA $=$ 67$^{\circ}$) with the sensitivity of 0.4 mJy beam$^{-1}$. Note that substructures are different between Band 3 and 7. (c) Spectral index ($\alpha$) map between Band 3 and 7. The $\alpha$ values are overall small with a mean value of 2.28 $\pm$ 0.02, and also they are distributed asymmetrically. {\rvs (d) Statistical error map of spectral indexes. The white contours of (c) and (d) mark where the error level is 0.07.}}
        \label{fig:fig2}
    \end{figure*}

    Figure \ref{fig:fig2}b shows a Band 7 continuum image of WL 17 with an angular resolution of 0.107$\arcsec$ $\times$ 0.104$\arcsec$.
    As mentioned in Section \ref{sec:observations},
    we used only the extended configuration archival data to focus on disk substructures (the second execution block in Table \ref{tab:tab1}).
    Compared with the Band 3 image shown in Figure \ref{fig:fig2}a, the Band 7 image reveals different substructures: an off-center hole and an asymmetric ring.
    These substructures are also consistent with those reported by \citet{2021ApJ...922..150G}.
    The hole has a radius of $\sim$0.04$\arcsec$ (5 au), and its center is shifted toward the southwest direction.
    The ring is asymmetric about the disk minor axis: specifically, the northeastern part has a larger width of $\sim$0.13$\arcsec$ (18 au) than the southwestern part with a width of $\sim$0.10$\arcsec$ (14 au).
    The ring also has an asymmetric brightness distribution along the azimuthal direction, showing that the maximum intensity is 22.7 mJy beam$^{-1}$ at PA = 60$^{\circ}$ while the minimum intensity is 16.6 mJy beam$^{-1}$ at PA = 180$^{\circ}$.
    The total flux within the 5$\sigma_{\rm B7}$-contour region with a radius of $\sim$0.23$\arcsec$ (32 au) is measured to be 125.81 $\pm$ 1.36 mJy.

    Likewise, we estimate the dust mass of the disk in Band 7.
    The dust opacity in Band 7 is calculated as 3.50 cm$^2$ g$^{-1}$ at a central frequency of 350 GHz by the equation in \citet{1990AJ.....99..924B}.
    With the same dust temperature and distance of 30 K and 137 pc, the dust mass in Band 7 is 13 $M_{\oplus}$, which is half of that estimated in Band 3.
    In other words, when assuming the typical dust opacity values ($\beta$ $=$ 1) computed in \citet{1990AJ.....99..924B}, $\kappa_{\nu}$ = 0.975 cm$^2$ g$^{-1}$ in Band 3 and $\kappa_{\nu}$ = 3.50 cm$^2$ g$^{-1}$ in Band 7, there is a discrepancy in the dust mass estimation between these two bands: 26 $M_{\oplus}$ and 13 $M_{\oplus}$.
    To match up these dust masses,
    the (sub-)mm dust opacity spectral index
    between Band 3 and 7 is lower than the typical value ($\beta$ = 1) employed in \citet{1990AJ.....99..924B}.
    Such a low dust opacity index in (sub-)mm wavelengths suggests the possible presence of mm/cm-sized large grains in the optically thin disk midplane \citep{1993Icar..106...20M, 2001ApJ...553..321D, 2006ApJ...636.1114D}.
    Grain growth in WL 17 will be further discussed through the $\beta$ analysis in the following subsection.

    \subsection{Spectral Index}
    \label{sec:spectral_index}

    A dust opacity ($\kappa_\nu$) is reasonably well described as a power-law function of frequency, $\kappa_\nu$ $\propto$ $\nu^\beta$, in (sub-)mm wavelengths \citep[e.g.,][]{1983QJRAS..24..267H, 1990AJ.....99..924B, 1991ApJ...381..250B, 1993Icar..106...20M}.
    Based on theoretical studies, the (sub-)mm dust opacity spectral index ($\beta$) depends on various properties of dust grains, such as size, shape, composition, and porosity \citep[e.g.,][]{1993Icar..106...20M, 1994ApJ...421..615P, 2001ApJ...553..321D, 2006ApJ...636.1114D, 2014A&A...568A..42K}.
    Among these dust properties, it is highly sensitive to the maximum grain size ($a_{\rm max}$): $\beta$ $\lesssim$ 1.0 at $\lambda$ $=$ 1 mm corresponds to $a_{\rm max}$ $\gtrsim$ 3 mm \citep{2006ApJ...636.1114D}.
    Thus, $\beta$ is commonly utilized to investigate grain growth in YSOs \citep[e.g.,][]{2009ApJ...696..841K}.

    The dust opacity index ($\beta$) is directly linked to the spectral index ($\alpha$) in (sub-)mm wavelengths.
    The spectral index is defined as $\alpha$ $=$ \rm{log}($I_{\nu_1}$/$I_{\nu_2}$) / \rm{log}($\nu_1$/$\nu_2$),
    where $I_{\nu_1}$ and $I_{\nu_2}$ are specific intensities at certain frequencies $\nu_1$ and $\nu_2$.
    The relationship between the spectral index and the dust opacity index is derived as the following equation \citep[e.g.,][]{2016ApJ...829L..35T, 2019MNRAS.486.3907P}:
    \begin{equation}
        \label{alpha1}
        \alpha = 3 - \frac{h \nu}{k_B T_{\rm dust}} \frac{e^{h \nu / k_B T_{\rm dust}}}{e^{h \nu / k_B T_{\rm dust}} - 1} + \frac{\tau_\nu}{e^{\tau_\nu} - 1} \beta,
    \end{equation}
    where $h$ is the Planck constant, $\nu$ is the geometric mean frequency between frequencies $\nu_1$ and $\nu_2$, $k_B$ is the Boltzmann constant, $T_{\rm dust}$ is the dust temperature, and $\tau_\nu$ is the geometric mean optical depth between optical depths $\tau_{\nu_1}$ and $\tau_{\nu_2}$.
    Note that for the optically thin case ($\tau_{\nu}$ $\ll$ 1), assuming the Rayleigh-Jeans approximation, Equation \ref{alpha1} can be simply expressed as
    $\alpha$ $=$ $\beta$ $+$ 2 \citep[e.g.,][]{2009ApJ...696..841K}.
    On the other hand, for the highly optically thick ($\tau_\nu$ $\gg$ 1) case in the Rayleigh-Jeans regime, Equation \ref{alpha1} is expressed
    as $\alpha$ $=$ 2, which means that the dust opacity index cannot be estimated from the spectral index at all.
    Thus, in order to investigate grain growth, it is necessary to first measure optical depth.
    
    Using the Planck function, we
    compute the optical depths for the Band 3 and 7 dust continuum emissions within the 5$\sigma$-contour regions {\rvs (Figures \ref{fig:fig2}a and \ref{fig:fig2}b)}.
    The mean intensities in Band 3 and 7 are {\rvs 0.707 mJy beam$^{-1}$} and 10.8 mJy beam$^{-1}$, respectively.
    We adopt a dust temperature of 30 K for this calculation because this value is considered the mean dust temperature of the ring based on the radiative transfer modeling introduced in Section \ref{sec:modeling_setup}.
    The mean optical depth values are calculated to be {\rvs 0.35} in Band 3 and 0.57 in Band 7, and then Equation \ref{alpha1} is derived as $\alpha$ $=$ 1.84 $+$ {\rvs 0.79$\beta$.}
    {\rvf The optical depths at peaks are {\rvs 0.72} in Band 3 and 2.40 in Band 7.}
    {\rvf We acknowledge that the small $\alpha$ could be caused by a combination of relatively high optical depths, temperature gradients, and/or self-scattering in the line of sight \citep[e.g.,][]{2017ApJ...840...72L, 2018ApJ...868...39G, 2021ApJ...923..270L, 2023arXiv230801972X}. However, we argue that it would be limited to small regions so our data may not be affected significantly. Therefore, with caution we consider that} 
    the emissions in the two bands are marginally optically thin, and we will discuss it further in Section \ref{sec:modeling_results}.
    Recently, \citet{2021ApJ...922..150G} also showed consistent results that the Band 3 and 7 continuum emissions are marginally optically thin in the entire disk region.
    We can thus estimate grain size from the spectral index between Band 3 and 7.

    Figure \ref{fig:fig2}c shows the spectral index map obtained from the Band 3 and 7 dust continuum images in Figures \ref{fig:fig2}a and \ref{fig:fig2}b.
    Only the intensity values above the 5$\sigma$ level are used to calculate the spectral index.
    This spectral index map has two interesting features.
    First, the index values are overall low with a mean value of 2.28 $\pm$ 0.02 in a narrow range of 2.01 and 2.71.
    The uncertainty of this mean value is determined from the uncertainties of the Band 3 and 7 total flux values measured in the previous subsections.
   Note that the absolute flux calibration uncertainties are $\sim$5$\%$ in Band 3 and $\sim$10$\%$ in Band 7, resulting in about 0.12 variations of the spectral index measurement ($\alpha$ $=$ 2.28$^{+0.11}_{-0.12}$), which implies that the mean spectral index is still low.
   {\rvf In addition, a variation of spectral indexes appears.
   {\rvs The white contours in Figure \ref{fig:fig2}c and \ref{fig:fig2}d mark where the statistical error of spectral indexes based on the sensitivities of both bands is 0.07. The inner region, which has a smaller error, shows a variation of spectral indexes up to $\Delta \alpha \sim 0.26$. This indicates that the variation of spectral indexes is not negligible, which is larger than $3\sigma$. Note that the spectral index error due to absolute flux uncertainties does not affect the spatial variation.}}
   
    Using the above $\alpha$ equation, the dust opacity index $\beta$ is calculated as {\rvs 0.56} $\pm$ 0.03.
    Several theoretical studies have shown $\beta$ profiles as a function of $a_{\rm max}$ and $q$ within a similar wavelength range to the interval between Band 3 and 7, where $a_{\rm max}$ and $q$ are the maximum grain size and index of the power-law grain size distribution
    $n(a)$ $\propto$ $a^{-q}$, respectively \citep[e.g.,][]{2001ApJ...553..321D, 2010A&A...512A..15R, 2018ApJ...869L..45B}.
    According to these profiles, $\beta_{\rm B3-B7}$ $=$ {\rvs 0.56} corresponds to $a_{\rm max}$ $=$ 0.2$-$20 cm and $q$ $=$ 2.5$-$3.0.
    Given the estimated age of this late Class I protostar \citep[$\lesssim$0.7 Myr;][]{2015ApJS..220...11D}, dust grains have already grown up to a few centimeters in size during the protostellar stages.
    Indeed, grain growth to mm/cm sizes on the protostellar disk scale, demonstrated by such a low $\beta$ value, has so far been reported in only a few Class I sources, such as EC 53 \citep[][]{2020ApJ...889...20L} and CB 26 \citep[][]{2021A&A...646A..18Z}.
    Second, the $\alpha$ values are asymmetrically distributed in the disk.
    It suggests that dust grains are differently distributed depending on their sizes.
\section{Modeling Analysis}
\label{sec:modeling_analysis}

As described in Sections \ref{sec:band3} and \ref{sec:band7}, the disk substructures are different between Band 3 and 7: a central hole and a symmetric ring in Band 3, while an off-center hole and an asymmetric ring in Band 7.
In Section \ref{sec:spectral_index}, from the intrinsic difference between the brightness distributions in these two bands, we obtain the asymmetric spectral index ($\alpha_{\rm mm}$) map with a low mean value of 2.28 $\pm$ 0.02, which implies rapid grain growth and dust segregation at the protostellar disk scale.
Thus, in this section, to verify these two features suggested by the spectral index map, we conduct radiative transfer modeling with the public code RADMC-3D \citep{2012ascl.soft02015D} and analyze the modeling results.

    \subsection{Modeling Setup}
    \label{sec:modeling_setup}

    The protostellar properties for our modeling analysis are based on previous studies.
    As mentioned in Section 1, WL 17 is an M3 protostar \citep{2010ApJS..188...75M}, whose typical effective temperature ($T_{\rm eff}$) is 3410 K \citep{2014ApJ...786...97H}.
    We adopt this temperature for our models.
    Note that \citet{2017ApJ...840L..12S} used a similar effective temperature of 3400 K, which was measured by
    Keck NIRSPEC observations in \citet{2005AJ....130.1145D}.
    From the Spitzer Space Telescope observations, \citet{2015ApJS..220...11D} obtained the extinction-corrected infrared spectral index ($\alpha_{\rm IR}^{'}$), bolometric temperature ($T_{\rm bol}^{'}$), and bolometric luminosity ($L_{\rm bol}^{'}$) of WL 17 as 0.72, 420 K, and 0.64 $L_{\odot}$, respectively.
    Based on $\alpha_{\rm IR}^{'}$ and $T_{\rm bol}^{'}$, the authors showed that WL 17 is in the late Class I stage, supporting previous results \citep[e.g.,][]{2009ApJ...692..973E, 2009ApJS..181..321E}.
    Also, the most probable duration of the Class 0$+$I stage was calculated to be from 0.46 to 0.72 Myr by comparing the populations between the Class I sample and the reference Class comprising all of the Class II sample and part of the Class III sample \citep{2015ApJS..220...11D}.
    In order to estimate the protostellar mass and luminosity of WL 17, we refer to the MIST isochrone, which covers a wide age range from 0.1 Myr to 20 Gyr \citep{2016ApJ...823..102C}.
    According to the isochrone, protostars with 3410 K and 0.46$-$0.72 Myr have 0.3 $M_{\odot}$ and 0.45$-$0.62 $L_{\odot}$.
    This protostellar luminosity range is consistent with the extinction-corrected bolometric luminosity.
    For these reasons, we assume the protostellar mass and luminosity as 0.3 $M_{\odot}$ and 0.5 $L_{\odot}$, respectively.
    Note that this protostellar luminosity is the same as that adopted in \citet{2017ApJ...840L..12S}.
    Regarding disk geometry, as described in Section \ref{sec:band3}, our estimates are consistent with the previous results.
    The inclination and position angles of the disk are thus fixed at 30$^{\circ}$ and 58$^{\circ}$, respectively, for the modeling analysis.


    \begin{figure} 
        \centering
        \includegraphics[scale=0.6]{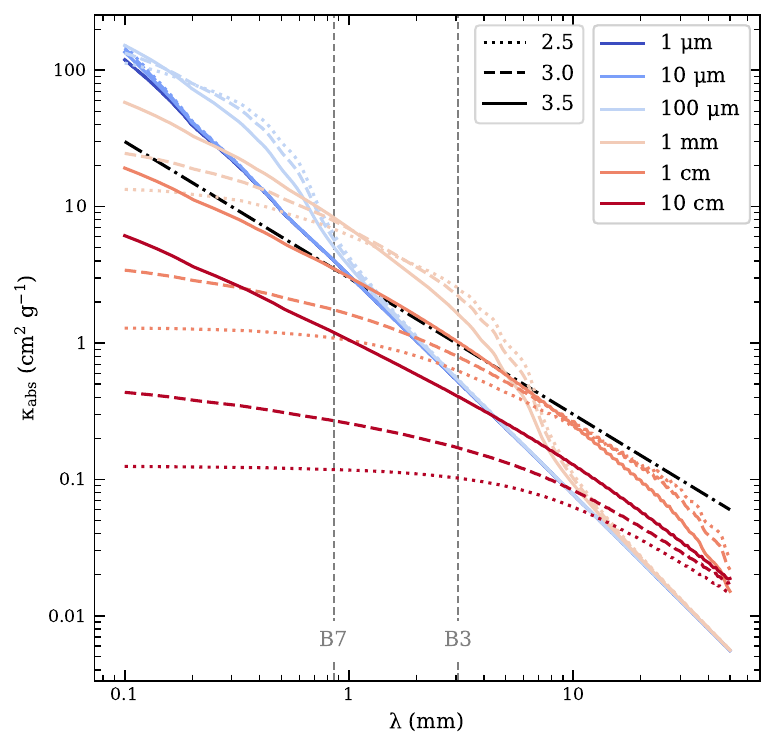}
        \caption{DIANA dust absorption opacities used for radiative transfer modeling. The {\rvf fifteen adopted} opacities have different line colors and styles depending on $a_{\rm max}$ and $q$. Two grey dashed vertical lines correspond to ALMA Band 3 (3.1 mm) and 7 (0.87 mm) wavelengths, respectively. For comparison, the opacity computed in \citet{1990AJ.....99..924B} is indicated by a black dash-dotted line. Note that this widely-used opacity is particularly similar to the DIANA opacity with $a_{\rm max}$ $=$ 1 cm and $q$ $=$ 3.5.}
        \label{fig:fig3}
    \end{figure}
    
    We employ dust opacities computed by the DIsc ANAlysis (DIANA) project \citep{2016A&A...586A.103W}.
    The opacities follow a power-law size distribution, $n(a)$ $\propto$ $a^{-q}$ from $a_{\rm min}$ = 0.05 $\mu$m to $a_{\rm max}$, where $q$ is the power-law index, $a_{\rm min}$ is the minimum grain size, and $a_{\rm max}$ is the maximum grain size.
    In order to constrain the size distribution of dust grains, we parameterize $a_{\rm max}$ and $q$: $a_{\rm max}$ $=$ \{10 $\mu$m, 100 $\mu$m, 1 mm, 1 cm, 10 cm\}, and $q$ $=$ \{2.5, 3.0, 3.5\}.
    The other parameters for constraining dust properties are assumed to be the same as those defined in \citet{2016A&A...586A.103W}.
    All the opacities used for our models are shown in Figure 3.
    Particularly, the opacity with $a_{\rm max}$ = 1 cm and $q$ = 3.5 is comparable to the widely-used one calculated by \citet{1990AJ.....99..924B}: $\kappa_{\nu}$ = 10 ($\nu$ / 1 THz)$^\beta$ and $\beta$ = 1.
    The opacity in \citet{1990AJ.....99..924B} is also supported by the opacities of mm/cm-sized large grains computed in several previous studies \citep[e.g.,][]{{2009ApJ...700.1502A},{2011ApJ...732...42A},{2018ApJ...869L..45B},{2019MNRAS.486.3907P}}.

    
    The spatial grid of our models is defined in spherical coordinates ($r$, $\theta$, $\phi$) that RADMC-3D supports \citep[e.g.,][]{2020A&A...633A.137D}, and we employ azimuthally symmetric models, i.e., independent of $\phi$.
    The $r$ grid is spaced logarithmically: it has a total of 60 cells and starts from a dust sublimation radius ($R_{\rm sub}$ $=$ 0.05 au) to 50 au, which is far enough to cover the entire disk region in the radial direction ($R_{\rm disk}$ = 22.7 au; \citealt{2017ApJ...840L..12S}).
    Note that $R_{\rm sub}$ is calculated to be 0.05 au by the following equation: $R_{\rm sub}$ $=$ ($L_*$ / (4$\pi \sigma_{\rm SB} T_{\rm sub}^4$))$^{0.5}$, where $L_*$ is the stellar luminosity, $\sigma_{\rm SB}$ is the Stefan-Boltzmann constant, and $T_{\rm sub}$ is the dust sublimation temperature.
    $L_*$ is adopted to be 0.5 $L_{\odot}$, as mentioned above, and we assume that $T_{\rm sub}$ is 1500 K \citep[e.g.,][]{2009ApJ...700.1502A}.
    To more specifically describe the disk substructures shown in Figure 1, we divide the $r$ grid into two parts: hole and ring regions.
    These two regions are separately sampled on a logarithmic scale.
    The hole region has 20 cells from 0.05 au to 8 au, and the ring region has 40 cells from 8 au to 50 au.
    The $\theta$ grid is spaced linearly: it has a total of 30 cells and starts from 75$^\circ$ to the disk midplane ($\theta$ = 90$^\circ$).
    We confirmed that this range is large enough to cover a few times the dust scale height used by \citet{2017ApJ...840L..12S} and the dust scale height adopted in our modeling.
    Likewise, in order to sample the entire $\theta$ grid at a higher resolution toward the disk midplane, we divide the $\theta$ grid into two parts: upper and lower layers.
    The upper layer has 5 cells from 75$^\circ$ to 80$^\circ$, and the lower layer has 25 cells from 80$^\circ$ to 90$^\circ$.
    Next, the cylindrical radius $R$ and the vertical height $z$ are defined as $R$ $=$ $r$sin($\theta$) and $z$ $=$ $r$cos($\theta$), respectively, to express the physical quantities of our models below, such as temperature and density.

    We set the dust temperature distribution in the WL 17 disk based on an empirical relation between the disk-midplane temperature and the optically thin limit. 
    Assuming dust grains are in radiative equilibrium with a central protostar,
    dust temperature in an optically thin region is expressed as the following power-law function
    (e.g., Equation 5 from \citealt{2009ApJ...696..841K}):
    $T_{\rm thin}(R)$ $=$ $T_{\rm sub}$ ($R$ / $R_{\rm sub}$)$^{-2 / (4+\beta)}$,
    where $T_{\rm sub}$ is the dust sublimation temperature, and $R_{\rm sub}$ is the sublimation radius at $T_{\rm dust}$ = $T_{\rm sub}$, and $\beta$ is the dust opacity index.
    Through a detailed radiative transfer modeling analysis, \citet{2015ApJ...808..102K} showed that
    the dust temperature distribution in the midplane of the FT Tau disk is roughly a third of such an optically-thin temperature distribution,
    particularly for distances ranging from a few au to tens of au from a central protostar,
    and the slope of the distribution is steeper in the midplane.
    Such a steeper slope is likely due to a higher optical depth in the midplane
    \citep[e.g.,][]{2003ApJ...592..255L, 2015ApJ...808..102K}.
    We apply these results to our models because the physical properties of both protostars and their surrounding disks are similar (e.g., $L_{\rm bol}$, $T_{\rm eff}$, and $R_{\rm disk}$; \citealt{2018ApJ...869...17L, 2019ApJ...882...49L}).
    The resultant radial dust temperature distribution that we adopt
    is
    defined as follows:
    \begin{equation}
        \label{eq:t_dust}
        T_{\rm mid}(R) = 30 \ \textrm{K} \left( \frac{R}{15 \ \textrm{au}} \right)^{-0.45}.
    \end{equation}
    We assume that the disk
    is vertically isothermal.
    Note that
    Equation \ref{eq:t_dust}
    is comparable with
    the midplane temperature
    obtained by \citet{2017ApJ...840L..12S} and that widely used for a flared disk in radiative equilibrium \citep[e.g.,][]{1997ApJ...490..368C, 1998ApJ...500..411D, 2001ApJ...560..957D}.

    We adopt Gaussian rings for the radial dust surface density distribution.
    Indeed, the Gaussian function has been often used to reproduce dust surface density distributions in protostellar and protoplanetary disks with rings and gaps \citep[e.g.,][]{2015PASJ...67..122M, 2018ApJ...869L..46D, 2020ApJ...891...48H}.
    For WL 17, \citet{2017ApJ...840L..12S} adopted a power-law dust surface density distribution to describe a typical protostellar system, consisting of a spherical envelope and an embedded disk.
    However,
    WL 17 has been classified as a late Class I protostar \citep[e.g.,][]{2009ApJ...692..973E, 2009ApJS..181..321E, 2015ApJS..220...11D},
    and \citet{2009A&A...498..167V} reported that the envelope has been nearly dissipated because there is no extended C$^{18}$O (3$-$2) emission within an angular radius of $\sim$40$\arcsec$.
    Furthermore, we add another Gaussian function to reproduce the weak emission in the center of the hole (hereafter called the inner disk), as shown in Figure 1.
    Thus, the radial dust surface density distribution for our models is defined as
    \begin{eqnarray}
    \label{eq:surfacedensity}
        \Sigma(R) = &\Sigma_{\rm hole}& \exp \left( -\frac{(R - R_{\rm hole})^2}{2{\sigma_{\rm hole}}^2} \right) +\\
                    &\Sigma_{\rm ring}& \exp \left( -\frac{(R - R_{\rm ring})^2}{2{\sigma_{\rm ring}}^2} \right),
    \end{eqnarray}
    where $\Sigma_{\rm hole}$ is the peak surface density of the inner disk at the radius $R_{\rm hole}$ ($=$ 0), $\sigma_{\rm hole}$ ($=$ 3.9 au) is the width of the inner disk, $\Sigma_{\rm ring}$ is the peak surface density of the ring at the radius $R_{\rm ring}$, and $\sigma_{\rm ring}$ is the width of the ring.
    Note that since the inner disk is not resolved at the current angular resolution, we simply assume that the inner disk is at the center of the hole and has the same width as the FWHM of the Band 3 synthesized beam: $R_{\rm hole}$ $=$ 0 au and $\sigma_{\rm hole}$ $=$ $\sigma_{\rm beam}$ $=$ 3.9 au.
    The other parameters, $\Sigma_{\rm hole}$, $\Sigma_{\rm ring}$, $R_{\rm ring}$, and $\sigma_{\rm ring}$, are set as free parameters for fitting.

    We adopt a power-law function for the radial dust scale height profile.
    Assuming that a disk is in hydrostatic equilibrium, this profile is determined from a power-law dust temperature distribution \citep[e.g.,][]{2015ApJ...808..102K}.
    Considering the dust-settling effect, we also adopt a new factor $f_{\rm H}$, which depends on the maximum grain size ($a_{\rm max}$).
    Specifically, based on the adopted $a_{\rm max}$ $=$ \{10 $\mu$m, 100 $\mu$m, 1 mm, 1 cm, 10 cm\}, the factor $f_{\rm H}$ is largely divided into three values.
    First, the models with $a_{\rm max}$ $=$ 10 $\mu$m have $f_{\rm H}$ = 1.
    Because $\mu$m-sized small grains are well mixed with gas, these grains are hardly settled down toward the disk midplane.
    Second, we assume $f_{\rm H}$ $=$ 0.5 for $a_{\rm max}$ $=$ 100 $\mu$m.
    Indeed, the dust scale height of a few hundred $\mu$m-sized grains is reported to be 0.1-0.8 times the gas scale height due to turbulence \citep[e.g.,][]{2019ApJ...886..103O, 2021ApJ...912..164D}, implying that such intermediate-sized grains are moderately mixed with gas.
    Last, the remaining models with $a_{\rm max}$ $\geq$ 1 mm have $f_{\rm H}$ = 0.1 because mm/cm-sized large grains are known to be highly settled down toward the disk midplane \citep[e.g.,][]{2011ApJ...732...42A, 2011ApJ...741....3K, 2016ApJ...816...25P, 2022ApJ...930...11V}.
    In summary, the radial dust scale height profile for our models is defined as
    \begin{equation}
        \label{eq:dust_scale_height}
        H(R) = f_{\rm H} H_0 \left( \frac{R}{R_0} \right)^h,
    \end{equation}
    where $f_{\rm H}$ is the dust settling factor, $H_0$ ($=$ 1.2 au) is the dust scale height at the radius $R_0$ ($=$ 15 au), and $h$ ($=$ 1.275) is the disk flaring index.
    Note that based on the dust temperature distribution (Equation \ref{eq:t_dust}), $H_0$ and $h$ are calculated as 1.2 au and 1.275, respectively.

    Finally, we search for model parameter sets reproducing the observed images best.
    This process includes two steps: finding parameter sets best fitting to the Band 3 image (Figure \ref{fig:fig1}) and comparing the model images generated using the parameters of the first step but at Band 7 with the observational Band 7 image (Figure \ref{fig:fig2}b).
    Also, to investigate grain properties, we select 15 pairs of ($a_{\rm max}$, $q$), covering wide ranges of the maximum grain size ($a_{\rm max}$) from 10 $\mu$m to 10 cm and the power-law size distribution index ($q$) from 2.5 to 3.5 (see also Figure \ref{fig:fig3}).
    For these 15 pairs of ($a_{\rm max}$, $q$), we individually set a model with four free parameters ($\Sigma_{\rm hole}$, $\Sigma_{\rm ring}$, $R_{\rm ring}$, and $\sigma_{\rm ring}$) and fit to the Band 3 data in the image domain.
    The best-fit model is obtained by maximizing the likelihood function, whose {\rvf logarithm} for each model produced by RADMC-3D on the image domain is defined by:
    \begin{eqnarray}
        {\rm ln}L &=& -\frac{1}{2} \Sigma \left[ \frac{(I_{\rm obsrv} - I_{\rm model})^2}{\sigma^2} + {\rm ln}(2 \pi \sigma^2) \right], \\
        \sigma^2 &=& \sigma_{\rm obsrv}^2 + (f I_{\rm model})^2,
    \end{eqnarray}
    where $I_{\rm obsrv}$ and $I_{\rm model}$ are the specific intensities of the Band 3 data and model, respectively, $\sigma_{\rm obsrv}$ $=$ 33 $\mu$Jy beam$^{-1}$ is the RMS noise level of the Band 3 data.
    {\rvf The $f$ parameter is introduced to consider unknown uncertainties in the fitting, which is caused by more complex distributions of quantities than assumed in our model: e.g., density, temperature, opacity, etc. \citep[e.g.,][]{2023arXiv230801972X}. {\rvs In circumstellar disks, such unknown perturbations are likely proportional to the intensity. Hence, we adopted $f I_{\rm model}$ as the unknown uncertainty. This form also has the advantage of giving more weight to the fitting of relatively faint emission, which is effective for evaluating the disk shape.}}
    Note that $f$ always converges to 0.1 regardless of initial free parameters; thus we fix it at 0.1 for this fitting.
    The maximizing process is performed through the Markov Chain Monte Carlo (MCMC) method using the public python package \texttt{emcee} \citep[][]{2013PASP..125..306F}.
    The uniform prior probability distributions for the free parameters are given over:  0 $<$ $\Sigma_{\rm hole}$ $<$ 10 g cm$^{-2}$, 0 $<$ $\Sigma_{\rm ring}$ $<$ 20 g cm$^{-2}$, 5 $<$ $R_{\rm ring}$ $<$ 30 au, and 0 $<$ $\sigma_{\rm ring}$ $<$ 10 au.
    The initial free parameters are sampled with 100 walkers and 1500 steps.
    The first 1000 steps are used to explore the parameter space for the burn-in phase.
    By adopting the medians of the burn-in phase as second initial values, the remaining 500 steps sample the posterior probability distribution.
    The best-fit parameters are taken as
    the medians of the final posterior probability distributions,
    and the uncertainties of the parameters are determined by the 68$\%$ confidence interval.
    In this way, the best-fit model in Band 3 is obtained for each of the 15 pairs with different grain size distributions 
    (Table 2).
    The reduced chi-square values ($\chi^2_{\rm reduced}$) for the 15 best-fit models are around 1.75, defined by
    \begin{equation}
        \chi^2_{\rm reduced} = \frac{\Sigma \left[ (I_{\rm obsrv} - I_{\rm model})^2 \ / \ \sigma^2_{\rm obsrv} \right]}{N - M},
    \end{equation}
    where $I_{\rm obsrv}$ and $I_{\rm model}$ are likewise the specific intensities of the Band 3 data and model, respectively, $\sigma_{\rm obsrv}$ $=$ 33 $\mu$Jy beam$^{-1}$ is the RMS noise level of the Band 3 data, $N$ is the number of pixels on the image plane, and $M$ is the number of free parameters.
    Finally, Band 7 model images are individually generated using the 15 best-fit parameter sets with different grain size distributions and are compared with the observed image. Based on the comparison, we constrain the size and spatial distributions of dust grains and identify the presence of additional substructures.

    \subsection{Modeling Results}
    \label{sec:modeling_results}

    \begin{figure*}
        \centering
        \includegraphics[scale=0.7]{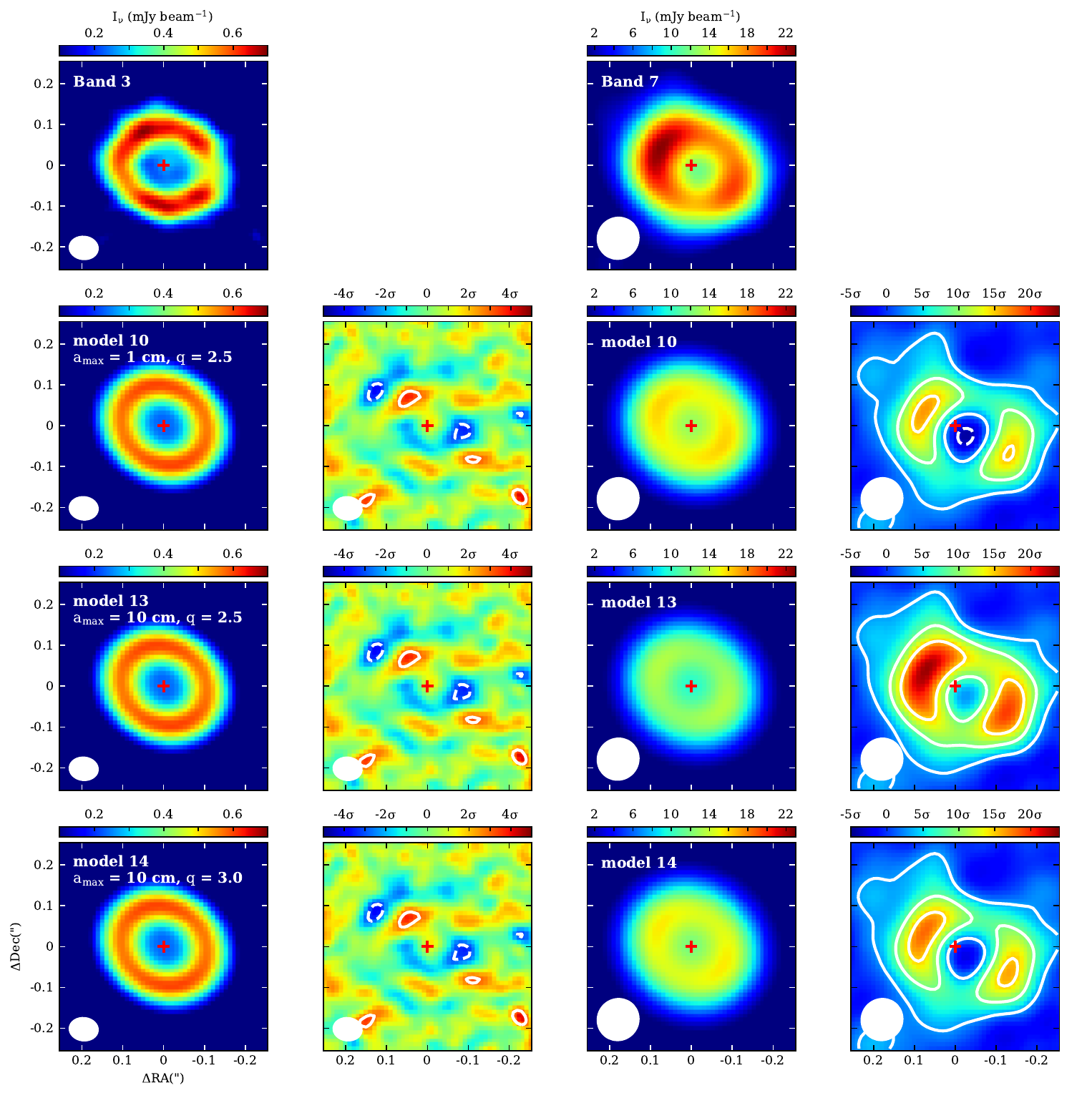}
        \caption{Part of the 15 best-fit models depending on $a_{\rm max}$ and $q$. The top two panels are the same as the Band 3 and 7 data shown in Figures \ref{fig:fig1} and \ref{fig:fig2}b. The red star indicates the position of the protostar. The remaining panels are the model images. From the left, the first shows the three Band 3 model images depending on $a_{\rm max}$ and $q$, the second column shows their residual maps {\rvf obtained by subtracting individual models from the Band 3 observational image}, the third column shows the Band 7 model images, and the last column shows their residual maps {\rvf obtained by subtracting individual models from the Band 7 observational image}. Contour levels in the Band 3 and 7 residual maps are \{$-$3, 3\}, where $\sigma_{\rm B3}$ $=$ 33 $\mu$Jy beam$^{-1}$, and \{$-$3, 3, 9, 15\}, where $\sigma_{\rm B7}$ $=$ 0.4 mJy beam$^{-1}$, respectively. Note that only these three models, mainly populated by cm-sized large grains, reproduce well the central hole in Band 7.}
        \label{fig:fig4}
    \end{figure*}

    \begin{figure*}
        \centering
        \includegraphics[scale=0.7]{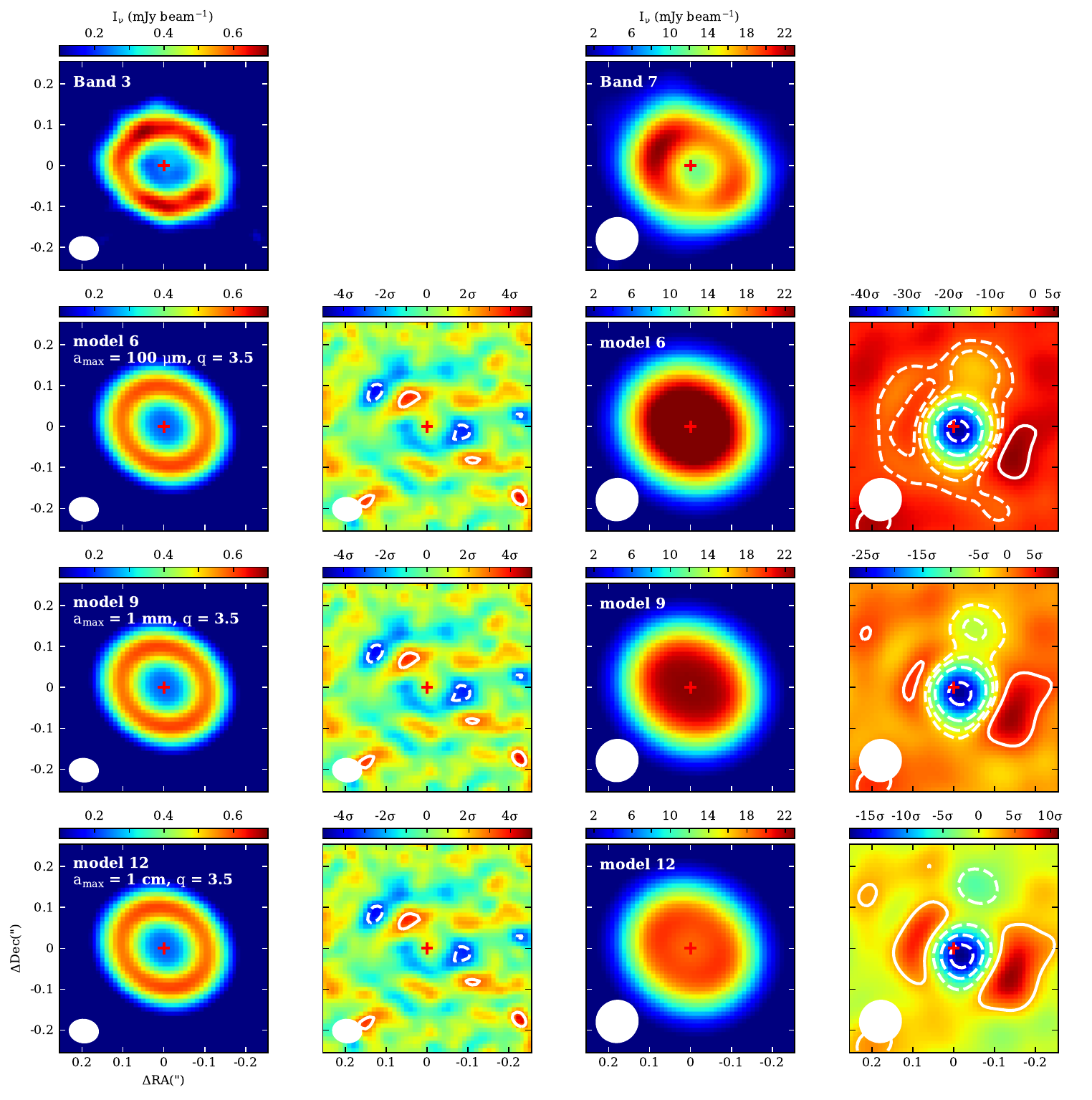}
        \caption{Part of the 15 best-fit models depending on $a_{\rm max}$ and $q$, same as Figure \ref{fig:fig4}, but different models. Contour levels in the Band 7 residual maps are \{$-$39, $-$24, $-$12, $-$6, $-$3, 3\} for model 6, \{$-$24, $-$12, $-$6, $-$3, 3\} for model 9, and \{$-$15, $-$9, $-$3, 3\} for model 12, $\sigma_{\rm B7}$ $=$ 0.4 mJy beam$^{-1}$. Note that these models do not reproduce the Band 7 data well, particularly the central hole.}
        \label{fig:fig5}
    \end{figure*}

    \begin{deluxetable*}{cccccccc}
        \label{tab:tab2}
        \tablenum{2}
        \tablecaption{Best-fit Parameters of the Adopted Disk Models}
        \tablehead{
            \colhead{Model} &
            \colhead{$a_{\rm max}$} &
            \colhead{$q$} &
            \colhead{$\Sigma_{\rm hole}$} &
            \colhead{$\Sigma_{\rm ring}$} &
            \colhead{R$_{\rm ring}$} &
            \colhead{$\sigma_{\rm ring}$} &
            \colhead{M$_{\rm dust}$} \\
            \colhead{} &
            \colhead{} &
            \colhead{} &
            \colhead{(g cm$^{-2}$)} &
            \colhead{(g cm$^{-2}$)} &
            \colhead{(au)} &
            \colhead{(au)} &
            \colhead{($M_{\earth}$)}}
        \startdata
            1 & 10 $\mu$m  & 2.5 & 0.31$^{+0.02}_{-0.02}$ &  3.24$^{+0.09}_{-0.10}$ & 15.96$^{+0.04}_{-0.04}$ & 3.09$^{+0.06}_{-0.06}$ & 95.2 \\ 
            2 &            & 3.0 & 0.31$^{+0.02}_{-0.02}$ &  3.24$^{+0.11}_{-0.10}$ & 15.96$^{+0.04}_{-0.04}$ & 3.09$^{+0.06}_{-0.06}$ & 95.2 \\ 
            3 &            & 3.5 & 0.31$^{+0.02}_{-0.02}$ &  3.24$^{+0.10}_{-0.10}$ & 15.96$^{+0.04}_{-0.03}$ & 3.09$^{+0.06}_{-0.06}$ & 95.2 \\ 
            \hline
            4 & 100 $\mu$m & 2.5 & 0.30$^{+0.02}_{-0.02}$ &  3.10$^{+0.11}_{-0.09}$ & 15.97$^{+0.04}_{-0.04}$ & 3.12$^{+0.06}_{-0.06}$ & 92.2 \\ 
            5 &            & 3.0 & 0.30$^{+0.02}_{-0.02}$ &  3.13$^{+0.10}_{-0.10}$ & 15.97$^{+0.04}_{-0.04}$ & 3.12$^{+0.06}_{-0.06}$ & 92.9 \\ 
            6 &            & 3.5 & 0.30$^{+0.02}_{-0.02}$ &  3.15$^{+0.09}_{-0.09}$ & 15.96$^{+0.04}_{-0.04}$ & 3.12$^{+0.06}_{-0.06}$ & 93.6 \\ 
            \hline
            7 & 1 mm       & 2.5 & 0.06$^{+0.00}_{-0.00}$ &  0.67$^{+0.02}_{-0.02}$ & 15.98$^{+0.04}_{-0.04}$ & 3.12$^{+0.06}_{-0.06}$ & 20.0 \\ 
            8 &            & 3.0 & 0.07$^{+0.00}_{-0.00}$ &  0.76$^{+0.02}_{-0.02}$ & 15.98$^{+0.04}_{-0.04}$ & 3.13$^{+0.06}_{-0.06}$ & 22.6 \\ 
            9 &            & 3.5 & 0.10$^{+0.01}_{-0.01}$ &  1.03$^{+0.03}_{-0.03}$ & 15.98$^{+0.04}_{-0.04}$ & 3.13$^{+0.06}_{-0.06}$ & 30.6 \\ 
            \hline
            10 & 1 cm       & 2.5 & 0.26$^{+0.02}_{-0.02}$ &  2.70$^{+0.09}_{-0.08}$ & 15.98$^{+0.04}_{-0.04}$ & 3.11$^{+0.06}_{-0.06}$ & 79.9 \\ 
            11 &            & 3.0 & 0.21$^{+0.01}_{-0.01}$ &  2.13$^{+0.07}_{-0.07}$ & 15.97$^{+0.04}_{-0.04}$ & 3.11$^{+0.06}_{-0.06}$ & 63.3 \\ 
            12 &            & 3.5 & 0.16$^{+0.01}_{-0.01}$ &  1.66$^{+0.05}_{-0.05}$ & 15.98$^{+0.04}_{-0.04}$ & 3.11$^{+0.06}_{-0.06}$ & 49.3 \\ 
            \hline
            13 & 10 cm      & 2.5 & 1.59$^{+0.09}_{-0.09}$ & 16.38$^{+0.43}_{-0.53}$ & 15.98$^{+0.04}_{-0.04}$ & 3.12$^{+0.06}_{-0.05}$ & 487.1 \\ 
            14 &            & 3.0 & 0.59$^{+0.06}_{-0.06}$ &  9.78$^{+0.33}_{-0.27}$ & 15.98$^{+0.04}_{-0.04}$ & 3.13$^{+0.06}_{-0.06}$ & 291.7 \\ 
            15 &            & 3.5 & 0.40$^{+0.02}_{-0.02}$ &  4.12$^{+0.14}_{-0.11}$ & 15.98$^{+0.04}_{-0.04}$ & 3.13$^{+0.06}_{-0.06}$ & 122.9 \\ 
        \enddata
        \tablecomments{These parameters are obtained by fitting the adopted models to the Band 3 data (Figure \ref{fig:fig1}): $\chi^2_{\rm reduced}$ $\simeq$ 1.75. The dust masses estimated from the dust continuum images in Band 3 (Figure \ref{fig:fig1}) and 7 (Figure \ref{fig:fig2}b) are 26 $M_{\earth}$ and 13 $M_{\earth}$, respectively.}
    \end{deluxetable*}

    All the 15 Band 3 models reproduce the Band 3 data well, regardless of the adopted dust opacities.
    Figures \ref{fig:fig4} and \ref{fig:fig5} show the Band 3 and 7 intensity and residual maps of part of the adopted models.
    The two left columns are the Band 3 intensity and residual maps.
    In the intensity maps, substructures are clearly seen, consisting of a large central hole ($R_{\rm ring}$ $\simeq$ 16 au) and a narrow ring ($\sigma_{\rm ring}$ $\simeq$ 3.1 au), which are consistent with the observed image in the left top panel.
    The residual maps show a noisy pattern mainly within the 3$\sigma_{\rm B3}$ level, indicative of good fittings.
    However, given that all the models explain the data well, we can see that there is a degeneracy between the opacity and the dust surface density: as listed in 
    Table 2,
    when the adopted opacity is lower, the best-fit surface density values ($\Sigma_{\rm hole}$ and $\Sigma_{\rm ring}$) become higher, and vice versa.
    This degeneracy is attributed to using only single-band data.


    When applying the physical parameters obtained from the Band 3 fittings to the Band 7 data, we find that only the models with cm-sized large grains reproduce the central substructure shown in Band 7.
    The right two columns in Figures \ref{fig:fig4} and \ref{fig:fig5} show the Band 7 intensity and residual maps.
    Compared with the Band 7 image in the top panel, the intensity maps in Figure \ref{fig:fig4} marginally show the substructures, whereas those in Figure \ref{fig:fig5} are highly saturated without the central hole.
    Additionally, the intensity maps of the remaining 9 models, which are not shown in Figure \ref{fig:fig5}, are also
    centrally peaked.
    The reason is that as shown in Figure \ref{fig:fig3}, these models have steeper $\beta$ slopes than cm-sized models in Figure \ref{fig:fig4}, resulting in a significant increase in intensity from Band 3 to 7.
    This difference is more evident in the residual maps.
    Figure \ref{fig:fig4} shows that
    the residuals of the three models with $a_{\rm max}$ $=$ 1 cm and $q$ $=$ 2.5, $a_{\rm max}$ $=$ 10 cm and $q$ $=$ 2.5, and $a_{\rm max}$ $=$ 10 cm and $q$ $=$ 3.0 are distributed mainly within the $\pm$3$\sigma_{\rm B7}$ level at the center.
    On the other hand, the residual maps in Figure \ref{fig:fig5} and of the remaining models show highly negative values up to $-$40$\sigma_{\rm B7}$ in the same central hole region.
    This difference indicates that cm-sized models reproduce the Band 7 data, particularly the central hole, relatively better than the other models.
    Furthermore, the optical depths of these three best-fit models are consistent with the observed mean values.
    For example, in the case of the model with $a_{\rm max}$ $=$ 10 cm and $q$ $=$ 2.5, the optical depths at the peak ($R$ $\simeq$ 16 au) are around 2.0 in both bands, while the other regions have a considerably low value of less than 0.5.
    Thus, these modeling results suggest that grain growth up to 1$-$10 cm in size has already occurred in the ring and inner disk.
    
    It appears that the three large-grain models do not reproduce the substructures observed in Band 7 fully yet.
    As shown in Figure \ref{fig:fig5}, there are asymmetrically distributed positive residuals in the ring region, implying the presence of dust grains less sensitive to the Band 3 wavelength.
    Indeed, the spectral index between residual intensities of Band 3 and 7 at the residual peak position of PA $=$ 60$^{\circ}$ is calculated to be 3.4$-$3.6, which is very close to that of ISM ($\alpha_{\rm ISM}$ $\simeq$ 3.7; e.g., \citealt{1999ApJ...524..867F, 2001ApJ...554..778L, 2014A&A...566A..55P, 2014A&A...571A..11P}).
    This suggests that the asymmetric residual regions have an addition of $\mu$m-sized small grains, which are presumably located in the upper layer of the disk.
    Note that these large-grain models have a smaller dust scale height for mm/cm-sized larger grains (a smaller $f_{\rm H}$ parameter; see Section \ref{sec:modeling_setup}).

    Furthermore, such grain growth up to a few centimeters in the ring can be supported by the azimuthal shift of the continuum peak observed in (sub-)mm wavelengths.
    {\rvf The Band 3 image has the strongest $3\sigma_{\rm B3}$ peak in the northeast from the center, although symmetric models fit it well overall. The peak of the Band 7 image is much stronger.}
    For example, \citet{2016MNRAS.458.3927B} performed two-dimensional hydrodynamic simulations to investigate the dynamics of dust grains in a transition disk with a narrow ring.
    The gas in the ring rapidly forms a horseshoe-shaped vortex driven by the Rossby-wave instability (RWI), which is similar to the ring shape in WL 17, and the distribution of dust grains along the vortex depends on the grain size: $\mu$m-sized small grains are in the vortex center, while cm-sized large grains are located ahead of the vortex.
    Particularly, considering gas self-gravity, the shift angle of 1$-$10 cm-sized grains is about 30$^{\circ}$ (see Figure 9 in \citealt{2016MNRAS.458.3927B}).

    \begin{figure}
        \centering
        \includegraphics[scale=0.7]{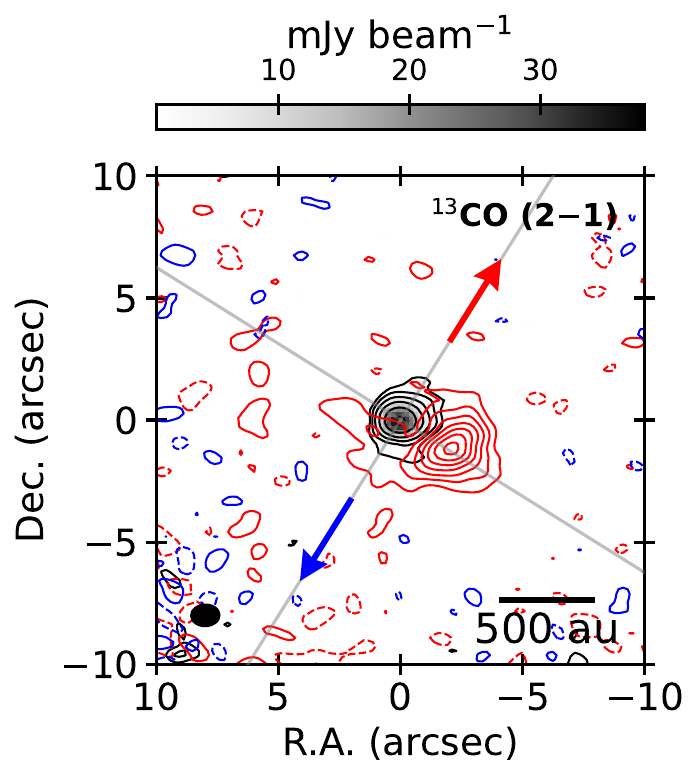}
        \caption{Moment 0 (integrated intensity) map of the redshifted and blueshifted $^{13}$CO (2$-$1) molecular line emission using the ALMA Band 6 ARI-L data. The redshifted emission (5.99$-$6.32 km s$^{-1}$) is shown as red contours. Its contour levels are \{$-$3, 3, 6, 9, ..., 21\} $\times$ $\sigma_{\rm ^{13}CO}$, where $\sigma_{\rm ^{13}CO}$ is 5.14 mJy beam$^{-1}$ km s $^{-1}$. The blueshifted emission (2.67$-$3.00 km s$^{-1}$) is shown as blue contours with levels of \{$-$3, 3\} $\times$ $\sigma_{\rm ^{13}CO}$. The size of the synthesized beam shown in the lower left is 1.255$\arcsec$ $\times$ 0.971$\arcsec$ (PA $=$ $-$88$^{\circ}$). Grey color scale and black contours denote the 1.3-mm continuum image of the same ARI-L data. Its contour levels are \{5, 10, 20, 40, 80, 160\} $\times$ $\sigma_{\rm B6}$, where $\sigma_{\rm B6}$ is 223.2 $\mu$Jy beam$^{-1}$. The synthesized beam size of the continuum is nearly the same as that of the $^{13}$CO (2$-$1) line emission. Red and blue arrows indicate redshifted and blueshifted outflows, respectively, and the systematic velocity of WL 17 is 4.5 km s$^{-1}$ \citep{2013A&A...556A..76V}. Grey lines indicate the position angles of the disk major and minor axes, 58$^{\circ}$ and 148$^{\circ}$, which are obtained from the ALMA Band 3 continuum image (Figure \ref{fig:fig1}). Note that the redshifted $^{13}$CO (2$-$1) emission is in the southwestern region along the disk major axis, while the blueshifted one is not seen.}
        \label{fig:fig6}
    \end{figure}

    To apply the result of dust distributions in a vortex to WL 17, we examine the disk rotation direction using the Additional Representative Images for Legacy (ARI-L; \citealt{2021PASP..133h5001M}) image of the ALMA Band 6 $^{13}$CO (2$-$1) molecular line emission (2019.1.01792.S; PI: Diego Mardones).
    Figure \ref{fig:fig6} shows the moment 0 (integrated intensity) map of the redshifted and blueshifted $^{13}$CO (2$-$1) emissions and the 1.3-mm dust continuum emission.
    Considering the angular resolution of the data, these emissions likely come from an envelope structure.
    The systematic velocity of this target is 4.5 km s$^{-1}$, which was obtained from previous JCMT HARP $^{12}$CO (3$-$2) observations \citep{2013A&A...556A..76V}.
    The redshifted component (5.99$-$6.32 km s$^{-1}$) is clearly seen in the southwestern region along the disk major axis.
    On the other hand, the blueshifted component (2.67$-$3.00 km s$^{-1}$) is not seen in the northeastern region, presumably due to the diffuse foreground material \citep[e.g.,][]{2009A&A...498..167V}.
    Indeed, a strong $^{12}$CO (3$-$2) self-absorption feature was identified in the blueshifted velocity range between 2$-$4 km s$^{-1}$ \citep[][]{2013A&A...556A..76V}.
    This moment map indicates that the envelope is rotating clockwise.
    We can thus expect that the disk is rotating clockwise as well.
    {\rvs Note that the near side of the disk is toward the northwest, and the far side toward the southeast. In the clockwise rotation, the Band 3 peak is ahead of the Band 7 peak.}
    As shown in Figures \ref{fig:fig2}a and \ref{fig:fig2}b, the shift angle between these two peaks is about 30$^{\circ}$, which corresponds to that of 1$-$10 cm-sized grains.
    In addition to our modeling analysis, grain growth to 10 cm may be supported by the continuum peak shift between Band 3 and 7.
    {\rvf }

\section{Discussion}
\label{sec:discussion}

As shown in Figures \ref{fig:fig1} and \ref{fig:fig2}, the protostellar disk surrounding WL 17 has a large central hole \citep[][]{2017ApJ...840L..12S, 2021ApJ...922..150G}.
In general, various scenarios for
such a
central
hole {\rvf have} so far been proposed: for example, grain growth \citep[e.g.,][]{2012A&A...544A..79B, 2021ApJ...907...80O}, photoevaporation \citep[e.g.,][]{2007MNRAS.375..500A, 2010MNRAS.401.1415O}, disk winds by magnetorotational instability \citep[e.g.,][]{2010ApJ...718.1289S}, and dynamical clearing by (sub-)stellar or planetary companions \citep[e.g.,][]{1994ApJ...421..651A, 2011ApJ...729...47Z, 2012A&A...545A..81P, 2018ApJ...864L..26B}.
To explain the hole in WL 17, some of these scenarios have been discussed in previous studies.
\citet{2017ApJ...840L..12S} suggested that photoevaporation is unlikely due to the high accretion rate expected in the Class I stage.
\citet{2018ApJ...865..102T} showed that disk winds can reproduce the hole, but their best-fit model cannot reproduce the inner disk, which is revealed in WL 17 (Figure \ref{fig:fig1}).
Last, \citet{2019AJ....158...41S} found that there is no stellar companion that can dynamically clear the hole, based on their radial velocity measurement.
As described in Section \ref{sec:modeling_results}, the azimuthal difference of the continuum peak between Band 3 and 7
also 
implies the possibility of planet formation \citep{2016MNRAS.458.3927B}.
In the following section, we discuss whether dynamical clearing by a planetary companion(s) (so-called planet-disk interaction) can explain the rapid grain growth and dust segregation identified by our modeling analysis.
In addition to the planet-disk interaction, given the early evolutionary stage, we discuss the possibility of protostellar infall for the origins of the features, as several hydrodynamic simulations have demonstrated that material infalling from an envelope onto a disk can form a dust ring and induce grain growth to mm/cm sizes within the ring \citep[e.g.,][]{2015ApJ...805...15B, 2022ApJ...928...92K}.

    \subsection{Planet-disk Interaction}

   \subsubsection{How do grains rapidly grow and segregate?}
    \label{sec:discussion_1}
    
    An interaction between a disk and a giant planet
    can explain the rapid grain growth occurring at the
    protostellar
    disk scale.
    \citet{2019ApJ...885...91D} performed two-dimensional hydrodynamic simulations to investigate how dust grains evolve by a single Jupiter-mass planet in a disk.
    These simulations assume
    that a disk around a 1-$M_{\odot}$ protostar has a radius of 34 au,
    and a 1-$M_J$ planet circularly orbits at 10 au for 4000 orbits (corresponding to $\sim$0.13 Myr).
    The simulations also consider
    dust coagulation and fragmentation.
    During the first 1000 orbits, the planet
    has already carved
    a gap in both gas and dust,
    and $\mu$m-sized small grains in the ring have rapidly grown up to cm-sized large grains.
    Also, after grains quickly reach a steady state within this first 1000 orbits, the size distribution of these grains does not change significantly during the remaining 3000 orbits.
    We emphasize that the initial conditions of the simulations are similar to WL 17, except for the protostellar mass, and the resulting size and spatial distributions of dust grains are highly comparable to those revealed by our modeling analysis (Section \ref{sec:modeling_results}).
    Assuming a smaller protostellar mass of 0.3 $M_{\odot}$, corresponding to the mass of WL 17 (Section \ref{sec:modeling_setup}), the first 1000 orbital period is calculated to be $\sim$57,700 yr, which is much less than the estimated age of the late Class I protostar \citep[$\lesssim$0.7 Myr;][]{2015ApJS..220...11D}.
    Thus, if there is already a single Jupiter-mass protoplanet orbiting WL 17, then this planet can rapidly carve a central large hole and a narrow ring and trigger grain growth in the ring during the Class I stage.
        
    A single Jupiter-mass planet can also explain grain growth in
    the inner disk of WL 17.
    \citet{2019ApJ...885...91D} demonstrated that, in addition to grain growth in the ring, during the first 1000 orbits, part of $\mu$m-sized small grains pass through the gap and then grow, resulting in an inner disk consisting of mm/cm-sized large grains.
    Previously, \citet{2012ApJ...755....6Z} also showed similar results:
    mm-sized grains can penetrate a gap carved by a 1-$M_J$ planet and form an inner disk, and $\mu$m-sized smaller grains can also penetrate the gap and grow rapidly to mm size in the inner disk.
    These simulation results are consistent with our modeling results that there are 1$-$10 cm-sized large grains in the inner disk of WL 17.

    The planet-disk interaction also
    interprets the dust segregation identified in
    the WL 17 disk.
    As shown in {\rvf Section \ref{sec:modeling_results},} along the ring in the azimuthal direction,
    cm-sized large grains are distributed symmetrically, whereas $\mu$m-sized small grains are distributed asymmetrically.
    \citet{2019ApJ...884L..41B} performed two-dimensional hydrodynamic simulations to examine how the radial and azimuthal distributions of dust grains are evolved by one or two Jupiter-mass planets in a protoplanetary disk.
    They assumed that a disk around a protostar of 0.85 $M_{\odot}$
    has a gas radius of 198 au, and that grains with various sizes ranging from 0.1 $\mu$m to 1 mm are annularly distributed between 50 and 100 au.
    They did not consider dust evolution, i.e., coagulation and fragmentation, but instead focused on changes in the spatial distribution of dust grains in the ring by the planet(s).
    Their simulations show that dust grains are quickly (less than 0.6 Myr) segregated in the radial and azimuthal directions by one or two Jupiter-mass planets.
    Particularly,
    when there is
    only a single planet with 5 $M_J$, mm-sized large grains, which are decoupled from gas, are concentrated at a pressure bump induced by the planet(s), while $\mu$m-sized small grains, well mixed with gas, are more widely distributed inside and outside the pressure bump.
    In addition to the radial direction, these large grains are symmetrically distributed along the azimuth, whereas the spatial distribution of the small grains is relatively more asymmetric, resulting in an off-center hole (see also Figure 1 in \citealt{2019ApJ...884L..41B}).
    This difference, {\rvf particularly in the azimuth,} is consistent with our modeling results, {\rvf although the radial difference cannot be investigated by our observational data due to limited angular resolutions compared to the ring width.}
    Also, \citet{2019ApJ...885...91D} showed
    similar results that a giant planet with 1 $M_J$ can rapidly cause dust segregation in a disk with a radius of 34 au, which is a much smaller disk than the above simulations ($R_{\rm disk}$ $=$ 198 au) but comparable to the disk around WL 17 ($R_{\rm disk}$ $\lesssim$ 20 au; Table 2).
    In the radial direction,
    as the grain size becomes larger, dust grains are more concentrated toward the ring induced by the planet (see also Figure 5 in \citealt{2019ApJ...885...91D}).
    {\rvf Although the radial dependence of dust distributions needs to be studied further with higher angular resolution data,} these two numerical simulations suggest that if there is already a Jovian planet of a mass $\lesssim$5 $M_J$ in the hole of the WL 17 disk, this planet can cause dust segregation as well as grain growth.

    As the second step, we estimate the mass of the
    putative
    planet in the
    central hole.
    \citet{2016PASJ...68...43K} proposed an empirical relationship between the observed gap width of a protoplanetary disk and the mass of a single giant planet in the gap.
    The formula is expressed as follows:
    \begin{equation}
        \frac{M_{\rm p}}{M_\star} = 2.1 \times 10^{-3} \left(\frac{\Delta_{\rm gap}}{R_{\rm p}}\right)^2 \left(\frac{h_{\rm p}}{0.05 R_{\rm p}}\right)^{3/2} \left(\frac{\alpha}{10^{-3}}\right)^{3/2},
    \end{equation}
    where $M_{\rm p}$ and $M_{\star}$ are the masses of the planet and the protostar, respectively, $\Delta_{\rm gap}$ is the gap width, $R_{\rm p}$ is the orbital radius of the planet, $h_{\rm p}$ is the gas scale height at $R_{\rm p}$, and $\alpha$ is the Shakura-Sunyaev viscosity parameter \citep{1973A&A....24..337S}.
    Note that \citet{2016PASJ...68...43K} defined the gap width $\Delta_{\rm gap}$ and the orbital radius $R_{\rm p}$ as $\Delta_{\rm gap}$ $=$ $R_{\rm out} - R_{\rm in}$ and $R_{\rm p}$ $=$ ($R_{\rm out} + R_{\rm in}$)/2, respectively, where $R_{\rm in}$ and $R_{\rm out}$ are the inner and outer edges of the gap.
    Assuming that the radial surface density profile of a ring is Gaussian,
    $R_{\rm in}$ and $R_{\rm out}$ are derived to be $R_{\rm in}$ $=$ $\sqrt{2\rm{ln}2}\sigma_{\rm in}$ and $R_{\rm out}$ $=$ $R_{\rm ring}$ $-$ $\sqrt{2\rm{ln}2}\sigma_{\rm ring}$, where $\sigma_{\rm in}$ and $\sigma_{\rm ring}$ are the {\rvf standard deviations} of the {\rvf Gaussian} inner disk and outer ring {\rvf (see also Equation \ref{eq:surfacedensity})}.
    Since $\sigma_{\rm in}$ is assumed to be 3.9 au (Section \ref{sec:modeling_setup}) and the best-fit parameter $\sigma_{\rm ring}$ is obtained as 3.1 au through the MCMC fitting
    (Table 2),
    $\Delta_{\rm gap}$ and $R_{\rm p}$ are calculated as 7.76 and 8.47 au, respectively.
    The gas scale height $h_{\rm p}$($R$ $=$ $R_{\rm p}$)
    is 0.58 au, which is derived from the dust scale height profile of our models without considering the dust settling effect (Equation \ref{eq:dust_scale_height}).
    Likewise, as adopted in our models (Section \ref{sec:modeling_setup}), the protostellar mass $M_{\rm s}$ is set to 0.3 $M_{\odot}$.
    Lastly, we assume that $\alpha$ is
    10$^{-3}$, which was used in the above previous theoretical simulations \citep[e.g.,][]{2016PASJ...68...43K, 2019ApJ...884L..41B, 2019ApJ...885...91D}.
    These input values provide a mass estimate of the potential planet in the hole
    to be $\sim$0.9 $M_J$.

    \subsubsection{How does a Jupiter-mass planet rapidly form?}
    \label{sec:discussion_2}

    Many theoretical studies have so far predicted that gravitational instability (GI) can
    form Jupiter-mass gas giants ($\lesssim$10 $M_{\rm J}$) in a massive protostellar disk around a 1-$M_{\odot}$ protostar within a few thousand years \citep[e.g.,][and references therein]{1997Sci...276.1836B, 1998ApJ...503..923B, 2002Sci...298.1756M, 2004ApJ...609.1045M, 2007prpl.conf..607D}.
    In particular, several of these theoretical studies have demonstrated the possibility of rapid giant planet formation by GI in a disk around an M-type protostar, which is the same spectral type as WL 17 \citep[e.g.,][]{2006ApJ...643..501B, 2016MNRAS.463.2480B, 2020A&A...633A.116M}.
    \citet{2006ApJ...643..501B} showed that within a few hundred years, a marginally gravitationally unstable disk ($R_{\rm disk}$ $=$ 20 au and $M_{\rm disk}$ $=$ 0.021$-$0.065 $M_{\odot}$) around an M-type protostar of 0.1 or 0.5 $M_{\odot}$ can develop spiral arms, and then a few Jupiter-mass clumps form in the spiral arms.
    \citet{2016MNRAS.463.2480B} showed that a gravitationally unstable disk ($R_{\rm disk}$ $\leq$ 30 au and $M_{\rm disk}$ $=$ 0.01$-$0.08 $M_{\odot}$) around an M-type protostar of 0.3 $M_{\odot}$ can develop spiral arms, and then these arms rapidly fragment into a number of dense clumps with an average mass of 0.3 $M_{\rm J}$.
    We emphasize that these two studies adopted disk radii, disk masses, protostellar masses, and midplane temperatures very similar to those of WL 17 (see also Section \ref{sec:modeling_setup}).
    Furthermore, \citet{2020A&A...633A.116M} showed that a gravitationally unstable and larger disk ($R_{\rm disk}$ $=$ 60$-$120 au and $M_{\rm disk}$ $=$ 0.040$-$0.083 $M_{\odot}$) around an M-type protostar of 0.2$-$0.4 $M_{\odot}$ can fragment and finally form Jupiter-mass protoplanets (2$-$6 $M_{\rm J}$) within a few thousand years.
    All these theoretical studies prove that GI can
    form Jupiter-mass protoplanets in a massive protostellar disk around an M-type protostar rapidly.

    We inspect whether WL 17 has a gravitationally unstable disk.
    In general, disk instability is determined by the Toomre $Q$ parameter \citep{1964ApJ...139.1217T}.
    The Toomre $Q$ parameter is defined as
    \begin{equation}
        Q = \frac{c_s \Omega}{\pi G \Sigma},
    \end{equation}
    where $c_s$ = $\sqrt{k_B T / \mu m_p}$ is the isothermal sound speed of an ideal gas, $\Omega$ = $\sqrt{G M_{\star} / R^3}$ is the Keplerian angular velocity, $G$ is the gravitational constant, and $\Sigma$ is the disk surface density.
    In the $c_s$ expression, $k_B$ is the Boltzmann constant, $T$ is the gas temperature, $\mu$ is the mean molecular weight, and $m_p$ is the proton mass.
    We assume that gas temperature is the same as dust temperature, hence $T$ follows the dust temperature distribution of the disk midplane (see Equation \ref{eq:t_dust}), and that $\mu$ is {\rvf 2.37} \citep[e.g.,][]{2008A&A...487..993K}.
    Also, the disk surface density $\Sigma$ can be derived from the best-fit radial dust surface density distribution $\Sigma_{\rm dust}$, which is obtained by the MCMC fitting (Section \ref{sec:modeling_results}), with a typical gas-to-dust ratio of 100.
    Note that protostellar disks around M-type protostars are gravitationally unstable when $Q_{\rm min}$ $\lesssim$ 0.9$-$1.5 \citep[e.g.,][]{2006ApJ...643..501B, 2016MNRAS.463.2480B}.

    \begin{figure*}
        \centering
        \includegraphics[scale=0.7]{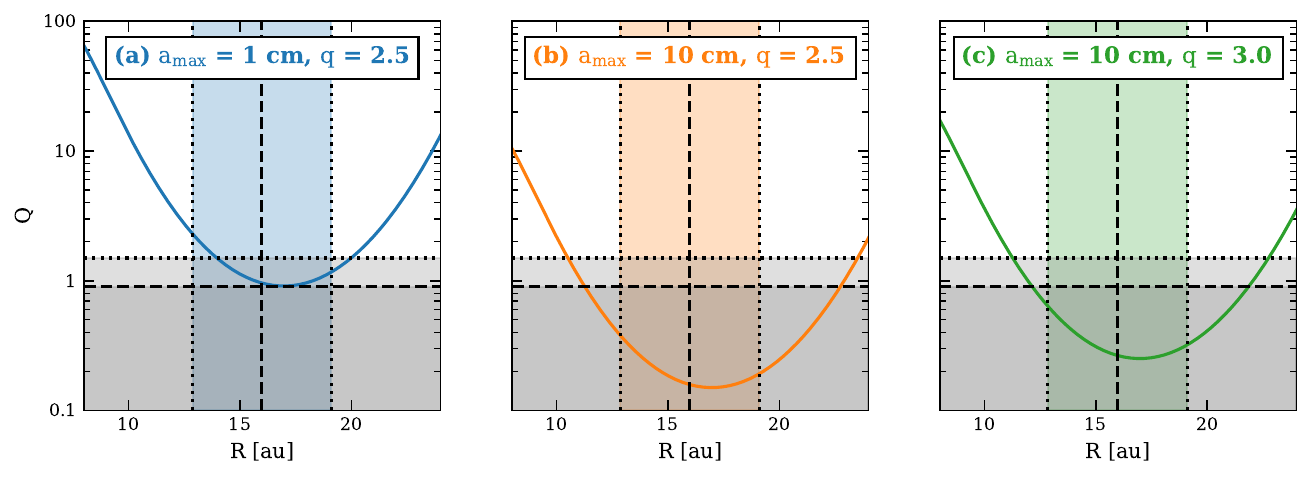}
        \caption{Toomre $Q$ parameter radial profiles of three best-fit models with (a) $a_{\rm max} = 1$ cm and $q = 2.5$, (b) $a_{\rm max} = 10$ cm and $q = 3.0$, (c) $a_{\rm max} = 10$ cm and $q = 3.0$. To focus on gravitational instability in the ring region, these profiles show between 8 and 24 au. In the vertical direction, the black dashed and dotted lines indicate $R_{\rm hole} \simeq 16.0$ au and $\sigma_{\rm ring} \simeq 3.1$ au of the best-fit models, and the ring region is shown as the shaded region in each panel. In the horizontal direction, the black dashed and dotted lines indicate the critical Toomre $Q$ parameter values of 0.9 \citep{2016MNRAS.463.2480B} and 1.5 \citep{2006ApJ...643..501B}, respectively, for a protostellar disk around an M-type protostar with 1 M$_{\odot}$. The grey shaded regions below these two lines are gravitationally unstable, meaning that Jupiter-mass planet(s) can rapidly form within the ring region.}
        \label{fig:fig7}
    \end{figure*}

    Figure \ref{fig:fig7} shows Toomre $Q$ parameter profiles as a function of radius between 8 and 24 au.
    These three profiles are obtained from the three best-fit models ($a_{\rm max}$ $=$ 1 cm and $q$ $=$ 2.5; $a_{\rm max}$ $=$ 10 cm and $q$ $=$ 2.5; and $a_{\rm max}$ $=$ 10 cm and $q$ $=$ 3.0), respectively.
    The Toomre $Q$ parameter profiles decrease toward the ring region and then increase again toward the outer region.
    Particularly, all the models have the lowest Toomre $Q$ parameters in the ring region: the model with $a_{\rm max}$ = 1 cm and $q$ = 2.5 has $Q_{\rm min}$ $\simeq$ 0.91, and the other two models with $a_{\rm max}$ = 10 cm and $q$ = 2.5 and with $a_{\rm max}$ = 10 cm and $q$ = 3.0 have $Q_{\rm min}$ $<$ 0.3.
    According to the above theoretical studies suggesting the condition needed for disk instability, $Q_{\rm min}$ $\lesssim$ 0.9$-$1.5 \citep[e.g.,][]{2006ApJ...643..501B, 2016MNRAS.463.2480B}, these low $Q$ values of all the three best-fit models indicate that WL 17 is gravitationally unstable in the ring region.
    Such instability supports the presence of a Jupiter-mass planet(s) as well as further planet formation in the future.
    In addition, assuming the typical gas-to-dust ratio of 100, the three models have high disk masses of 0.024 $M_{\odot}$, 0.146 $M_{\odot}$, and 0.088 $M_{\odot}$, respectively.
    These high disk masses are consistent with previous ALMA disk survey results that the dust mass of WL 17 is within the top 30$\%$ of the Class I protostellar disks in the $\rho$ Ophiuchus molecular cloud \citep{2019ApJ...875L...9W, 2019ApJS..245....2S, 2021ApJ...913..149E}.
    WL 17 is thus a gravitationally unstable and massive disk so that Jupiter-mass protoplanets have likely formed within a short time.
    If a protoplanet were detected, it would be the youngest one compared to the cases confirmed so far:
    a Jupiter-mass planet around AS 209 \citep[1$-$2 Myr;][]{2022ApJ...934L..20B},
    a Jupiter-mass planet around AB Aur \citep[4 Myr;][]{2022NatAs.tmp...76C}, two Jupiter-mass planets around PDS 70 \citep[5 Myr;][]{2019ApJ...879L..25I, 2021ApJ...916L...2B}, {\rvf and two Jupiter-mass planets around HD 163296 \citep[6 Myr; e.g.,][]{2021ApJS..257...18T, 2023A&A...674A.113I}.}
    {\rvf Note that the planets orbiting AS 209 and HD 163296 are kinematically identified by localized velocity perturbations in molecular line emission.}

    \subsection{Protostellar infall}

    An alternative to the planet-disk interaction is the protostellar infall scenario:
    material infalling from an envelope onto a disk can induce substructure and grain growth.
    \citet{2015ApJ...805...15B} showed that isotropic infall triggers the Rossby-wave instability (RWI), and this instability forms vortices, which can efficiently trap dust grains, particularly cm-sized large grains, and enhance grain growth.
    Recently, \citet{2022ApJ...928...92K} examined more realistic cases with anisotropic infall.
    It corresponds to filamentary inflows, called streamers, that have been lately reported both in (sub-)mm observations \citep[e.g.,][]{2014ApJ...793....1Y, 2019ApJ...880...69Y, 2020NatAs...4.1158P, 2022ApJ...925...32T, 2022A&A...667A..12V} and in numerical simulations \citep[e.g.,][]{2015MNRAS.446.2776S, 2017ApJ...846....7K, 2019A&A...628A.112K}.
    Their simulations show that streamers also trigger the RWI, forming vortices and
    pressure bumps in a disk.
    Furthermore, in these cases, dust grains drifting inward are concentrated on the annular pressure bumps and rapidly grow therein.
    These radial drift and grain growth result
    in a compact disk with a mean radius of 55 au and a ring structure consisting of mm/cm-sized large grains.
    This radius is in good agreement with the observational result that the mean dust disk radii of the Class 0/I sources are less than 50 au in the VLA/ALMA Nascent Disk and Multiplicity (VANDAM) survey of the Orion Molecular Clouds \citep{2020ApJ...890..130T, 2022ApJ...929...76S}.


    The infall scenario can be applied to WL 17, {\rvf although there is no feature of streamers found yet.}
    As described in Section \ref{sec:modeling_setup}, it is in the late Class I stage, and its age is estimated to be less than 0.72 Myr \citep{2015ApJS..220...11D}.
    Considering this young age, an envelope is expected to still remain, and \citet{2017ApJ...840L..12S} also suggested that this target is embedded in the remnants of its envelope.
    Indeed, the $^{13}$CO (2$-$1) emission in Figure \ref{fig:fig6}
    reveals
    that part of the inner envelope is rotating, although the outer envelope has almost dissipated \citep{2009A&A...498..167V}.
    Also, the dust continuum images in Band 3 and 7 (Figure \ref{fig:fig2}) show that the disk is compact but annularly structured, with a ring radius of 16 au (Table 2).
    In addition to the ring structures, the peak-intensity positions between the two bands differ by about 30$\arcdeg$ in the azimuthal direction.
    As mentioned in Section \ref{sec:modeling_results}, this difference suggests the presence of a vortex triggered by the RWI.
    These features are consistent with the above theoretical prediction,
    indicating
    that material is still being accreted from the surrounding envelope onto the disk.
    Thus, putative infall motion can also interpret the observed ring and the grain growth within the ring.

    However,
    this infall scenario cannot explain the grain growth in the inner disk of WL 17.
    The inner disk is indeed detected in the higher-resolution Band 3 image (Figure \ref{fig:fig1}), and our three best-fit models suggest that there are 1$-$10 cm-sized large grains in this inner disk.
    \citet{2015ApJ...805...15B} showed that since a vortex driven by infall efficiently traps 1$-$10 cm-sized dust grains, an inner disk is depleted of dust and contains only gas.
    In contrast, as discussed in Section \ref{sec:discussion_1}, the planet-disk interaction can explain the grain growth in the inner disk, thus making it the preferential scenario.

\section{Conclusions}
\label{sec:conclusions}

We used ALMA Band 3 and 7 archival data of the Class I protostellar disk WL 17.
Using these multi-wavelength and multi-configuration data, we present the Band 3 and 7 dust continuum images with angular resolutions of 0.07$\arcsec$ (10 au) and 0.1$\arcsec$ (14 au), respectively.
We also obtain a two-dimensional spectral index ($\alpha$) map between these two bands.
In addition, to further constrain grain properties, we perform radiative transfer modeling by testing several dust opacity models, which follow the power-law size distribution $n(a)$ $\propto$ $a^{-q}$ from $a_{\rm min}$ = 0.05 $\mu$m to $a_{\rm max}$\textbf, and compare the models with the multi-wavelength data.
The main results are summarized as follows:

\begin{enumerate}
    \item
    Disk substructures are clearly resolved in both Band 3 and 7, but they are significantly different:
    the Band 3 image shows a central hole and a symmetric ring, whereas the Band 7 image shows an off-centered hole and an asymmetric ring.
    These substructures are consistent with those reported by \citet{2017ApJ...840L..12S} and \citet{2021ApJ...922..150G}.

    \item
    The spectral index ($\alpha$) map between Band 3 and 7 shows two features: the $\alpha_{\rm mm}$ values are overall low with an average value of 2.28,
    and they are asymmetrically distributed.
    Based on the mean specific intensity, the WL 17 disk is estimated to be moderately optically thin in Band 3 and 7 ($\tau_{\nu}$ $\lesssim$ 0.6), indicating that the spectral index can be understood by grain sizes \citep[e.g.,][]{2006ApJ...636.1114D}.
    The spectral index map, therefore, suggests that (1) grains have already grown to mm/cm sizes, and (2) they are differently distributed depending on grain sizes.

    \item
    Only the models having a small scale height of dust grains and being populated by cm-sized large grains ($a_{\rm max}$ = 1 cm and $q$ = 2.5, $a_{\rm max}$ = 10 cm and $q$ = 2.5, and $a_{\rm max}$ = 10 cm and $q$ = 3.0) can explain the disk substructures, particularly the central holes observed both in Band 3 and 7.
    These modeling results suggest that grains have rapidly grown up to 1$-$10 cm in size and have been settled down toward the midplane during the protostellar phase.

    \item
    Nevertheless, the best models cannot fully explain the ring emission.
    Notably, in Band 7, the ring region has highly positive and asymmetric residuals.
    This
    can be interpreted as another $\mu$m-sized dust population, presumably in the upper layer, less sensitive to the Band 3 wavelength.
    It implies that cm-sized large grains in the midplane are symmetrically distributed {\rvf in the azimuthal direction,} whereas $\mu$m-sized small grains are asymmetrically distributed.
    
    \item
    The rapid grain growth and dust segregation identified by the modeling analysis can be explained by a single Jupiter-mass planet
    based on
    previous hydrodynamic simulations with a similar environment to WL 17.
    The high disk masses ($M_{\rm disk}$) of 0.024 $M_{\odot}$, 0.146 $M_{\odot}$, and 0.088 $M_{\odot}$ inferred from the three best models ($a_{\rm max}$ = 1 cm and $q$ = 2.5, $a_{\rm max}$ = 10 cm and $q$ = 2.5, and $a_{\rm max}$ = 10 cm and $q$ = 3.0) result in low minimum Toomre $Q$ parameter ($Q_{\rm min}$) values of 0.91, 0.15, and 0.25 in the ring region.
    It means that the disk is gravitationally unstable, so a giant planet has likely formed by gravitational instability (GI) during the Class I stage.
\end{enumerate}

\begin{acknowledgments}
    We are grateful to the anonymous referee for thoughtful comments.
    I.H. thanks Thiem Hoang, Chang Won Lee, Sang-Sung Lee, and Aran Lyo for helpful discussions.
    W.K. is supported by the National Research Foundation of Korea (NRF) grant funded by the Korea government (MSIT) (NRF-2021R1F1A1061794).
    This paper makes use of the following ALMA data: ADS/JAO.ALMA\#2015.1.00761.S, ADS/JAO.ALMA\#2019.1.01792.S.
    ALMA is a partnership of ESO (representing its member states), NSF (USA) and NINS (Japan), together with NRC (Canada), MOST and ASIAA (Taiwan), and KASI (Republic of Korea), in cooperation with the Republic of Chile.
    The Joint ALMA Observatory is operated by ESO, AUI/NRAO and NAOJ.
\end{acknowledgments}

%

\vspace{5mm}
\facilities{ALMA}


\software{CASA \citep{2007ASPC..376..127M}, RADMC-3D \citep{2012ascl.soft02015D}, emcee \citep{2013PASP..125..306F}}




\bibliography{WL17}{}
\bibliographystyle{aasjournal}












\end{document}